\documentclass[sn-mathphys-num,iicol]{sn-jnl}

\usepackage{graphicx}%
\usepackage{multirow}%
\usepackage{amsmath,amssymb,amsfonts}%
\usepackage{amsthm}%
\usepackage{mathrsfs}%
\usepackage[title]{appendix}%
\usepackage{xcolor}%
\usepackage{textcomp}%
\usepackage{manyfoot}%
\usepackage{booktabs}%
\usepackage{algorithm}%
\usepackage{algorithmicx}%
\usepackage{algpseudocode}%
\usepackage{listings}%

\usepackage{siunitx}
\usepackage{subcaption}
\usepackage{comment}
\usepackage{lineno}
\usepackage{lmodern}

\begin{document}
\title[Article Title]{Charge calibration of MALTA2, a radiation hard depleted monolithic active pixel sensor}
\author*[1,2]{\fnm{Lucian} \sur{Fasselt}}\email{lucian.fasselt@desy.de}
\author[3]{\fnm{Ignacio} \sur{Asensi Tortajada}}
\author[4]{\fnm{Prafulla} \sur{Behera}}
\author[1,2]{\fnm{Dumitru Vlad} \sur{Berlea}}
\author[5]{\fnm{Daniela} \sur{Bortoletto}}
\author[6]{\fnm{Craig} \sur{Buttar}}
\author[7]{\fnm{Valerio} \sur{Dao}}
\author[4]{\fnm{Ganapati} \sur{Dash}}
\author[3]{\fnm{Leyre} \sur{Flores Sanz de Acedo}}
\author[5]{\fnm{Martin} \sur{Gazi}}
\author[8]{\fnm{Laura} \sur{Gonella}}
\author[9]{\fnm{Vicente} \sur{González}}
\author[3]{\fnm{Sebastian} \sur{Haberl}}
\author[3]{\fnm{Tomohiro} \sur{Inada}}
\author[4]{\fnm{Pranati} \sur{Jana}}
\author[10]{\fnm{Long} \sur{Li}}
\author[3]{\fnm{Heinz} \sur{Pernegger}}
\author[3]{\fnm{Petra} \sur{Riedler}}
\author[3]{\fnm{Walter} \sur{Snoeys}}
\author[3]{\fnm{Carlos} \sur{Solans Sánchez}}
\author[3]{\fnm{Milou} \sur{van Rijnbach}}
\author[3,9]{\fnm{Marcos} \sur{Vázquez Núñez}}
\author[4]{\fnm{Anusree} \sur{Vijay}}
\author[3]{\fnm{Julian} \sur{Weick}}
\author[1,2]{\fnm{Steven} \sur{Worm}}

\affil*[1]{\orgname{Deutsches Elektronen-Synchrotron DESY}, \orgaddress{\city{Zeuthen}, \country{Germany}}}
\affil[2]{\orgname{Humboldt University of Berlin}, \orgaddress{\city{Berlin}, \country{Germany}}}
\affil[3]{\orgname{CERN}, \orgaddress{\city{Geneva}, \country{Switzerland}}}
\affil[4]{\orgname{Indian Institute of Technology Madras}, \orgaddress{\city{Chennai}, \country{India}}}
\affil[5]{\orgname{University of Oxford}, \orgaddress{\city{Oxford}, \country{UK}}}
\affil[6]{\orgname{University of Glasgow}, \orgaddress{\city{Glasgow}, \country{UK}}}
\affil[7]{\orgname{Stony Brook University}, \orgaddress{\city{New Yorck}, \country{US}}}
\affil[8]{\orgname{Università degli Studi di Trieste}, \orgaddress{\city{Trieste}, \country{Italy}}}
\affil[9]{\orgname{Universitat de València}, \orgaddress{\city{Valencia}, \country{Spain}}}
\affil[10]{\orgname{University of Birmingham}, \orgaddress{\city{Birmingham}, \country{UK}}} 

\abstract{MALTA2 is a depleted monolithic active pixel sensor (DMAPS) designed for tracking at high rates and typically low detection threshold of $\sim150\,\mathrm{e^-}$.
A precise knowledge of the threshold is crucial to understanding the charge collection in the pixel and specifying the environment for sensor application.
A simple procedure is developed to calibrate the threshold to unit electrons making use of a dedicated charge injection circuit and an Fe-55 source with dominant charge deposition of $1600\, \mathrm{e^-}$.
The injection voltage is determined which corresponds to the injection under Fe-55 exposure and is the basis for charge calibration.
The charge injection circuit incorporates a capacitance with design value of $\mathrm{C_{inj}}= \SI{230}{\atto\farad}$.
Experimentally, the average capacitance value for non-irradiated samples is found to be $\mathrm{C_{inj,exp}}= \SI{257}{\atto\farad}$.
The \SI{12}{\percent} divergence motivates the need for the presented precise calibration procedure, which is proposed to be performed for each MALTA2 sensor.
}

\keywords{DMAPS, Charge calibration, Tracking, High-energy physics}

\maketitle
\section{Introduction}
\label{sec:Intro}
\begin{figure}
\centering
\includegraphics[width=\linewidth]{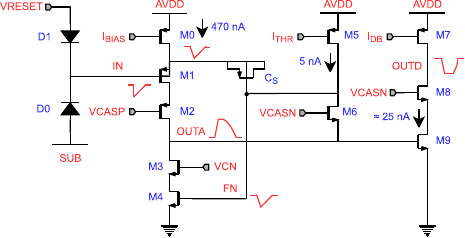}
\caption{MALTA2 front-end schematic including amplification, shaping and digitization of the analog signal per pixel \cite{Malta2_PiroFr}. 
IBIAS is the main biasing current and accounts for the majority of the power consumption. 
The current ITHR defines the speed of the feedback loop and is designed to linearly effect the threshold of the discriminator.}
\label{fig:MALTA2_FE}
\end{figure}

\begin{figure}
\centering
\includegraphics[width=\linewidth]{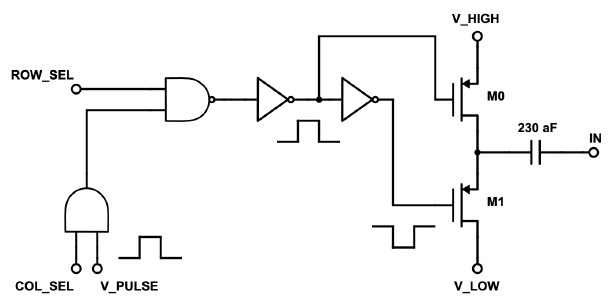}
\caption{MALTA2 charge injection circuit \cite{Berdalovic:2702884}.}
\label{fig:InjectionCircuit}
\end{figure}

MALTA2 is the second prototype of the MALTA family of depleted monolithic active pixel sensors (MAPS) designed in Tower \SI{180}{\nano\meter} CMOS imaging sensor technology \cite{Pernegger_2017,Berdalovic:2018tce,Malta2_PiroFr}. 
The MALTA2 pixel with a pitch of \SI{36.4}{\micro\meter} consists of either high resistivity epitaxial or Czochralski silicon. 
The front-end in every pixel, as shown in figure \ref{fig:MALTA2_FE}, is optimized for threshold settings down to $\sim150\,\mathrm{e^-}$ for detection efficiencies above $99\%$ as demonstrated for charged hadron beams \cite{vanRijnbach:2023gtl}.
The threshold is set globally by the different transistors, M0 to M9, by DAC settings corresponding to the gate of each transistor.
Every pixel is also equipped with a dedicated charge injection circuit (figure \ref{fig:InjectionCircuit}).
It makes use of two voltage settings $\mathrm{V_{HIGH}}$ and $\mathrm{V_{LOW}}$ and a $\mathrm{V_{PULSE}}$ signal. 
$\mathrm{V_{HIGH}}$ and $\mathrm{V_{LOW}}$ are subtracted by switching transistor M0 off and M1 on at the rising edge of $\mathrm{V_{PULSE}}$.
The resulting voltage difference is capacitively coupled to the input node of the front-end through a metal-to-metal connection that has a capacitance extracted from simulation of  $\mathrm{C_{inj}}= \SI{230}{\atto\farad}$ \cite{Berdalovic:2702884}.
The injected charge is therefore:
\begin{equation}
    \mathrm{Q_{inj}} = \mathrm{C_{inj}} \; \Delta \mathrm{V} =  \mathrm{C_{inj}} (\mathrm{V_{HIGH}}-\mathrm{V_{LOW}}).
\label{eq:QInjection}
\end{equation}

This document describes the procedure to measure the capacitance of the charge injection circuit of MALTA2 assuming a linear behaviour of the injected charge with respect to the difference of the $\mathrm{V_{HIGH}}$ and $\mathrm{V_{LOW}}$ voltages.
$\Delta \mathrm{V_{Fe55}}$ is the voltage difference that injects the same signal as an Fe-55 source.
It is assumed that the charge deposited by the \SI{5.9}{\kilo\electronvolt} $\mathrm{K_{\alpha}}$-line of an Fe-55 source is $1600\, \mathrm{e^-}$. 
Radioactive sources with well defined X-ray lines are commonly selected for silicon sensors as calibration reference because of its point-like charge deposition. 
For thin sensors such as MALTA2, Fe-55 is selected due to its relatively low peak energy compared to other gamma sources and X-ray fluorescence lines \cite{ThresholdMethod_Reconstr, 10.3389/fphy.2023.1231336}.
An alternative setup relying on Compton scattering requires multiple detectors as well as angle variation and is not suitable for frequent calibration of numerous samples \cite{mccormack2020newmethodsiliconsensor}.
In the following, the injected signals are quantified by the digital amplitude of the signal obtained through dedicated threshold scans, and $\Delta \mathrm{V_{Fe55}}$ will be measured by dedicated source scans.

\section{Methods}
\label{sec:Methods}
\subsection{Voltage measurement}
\begin{figure}
\centering
\includegraphics[width=\linewidth]{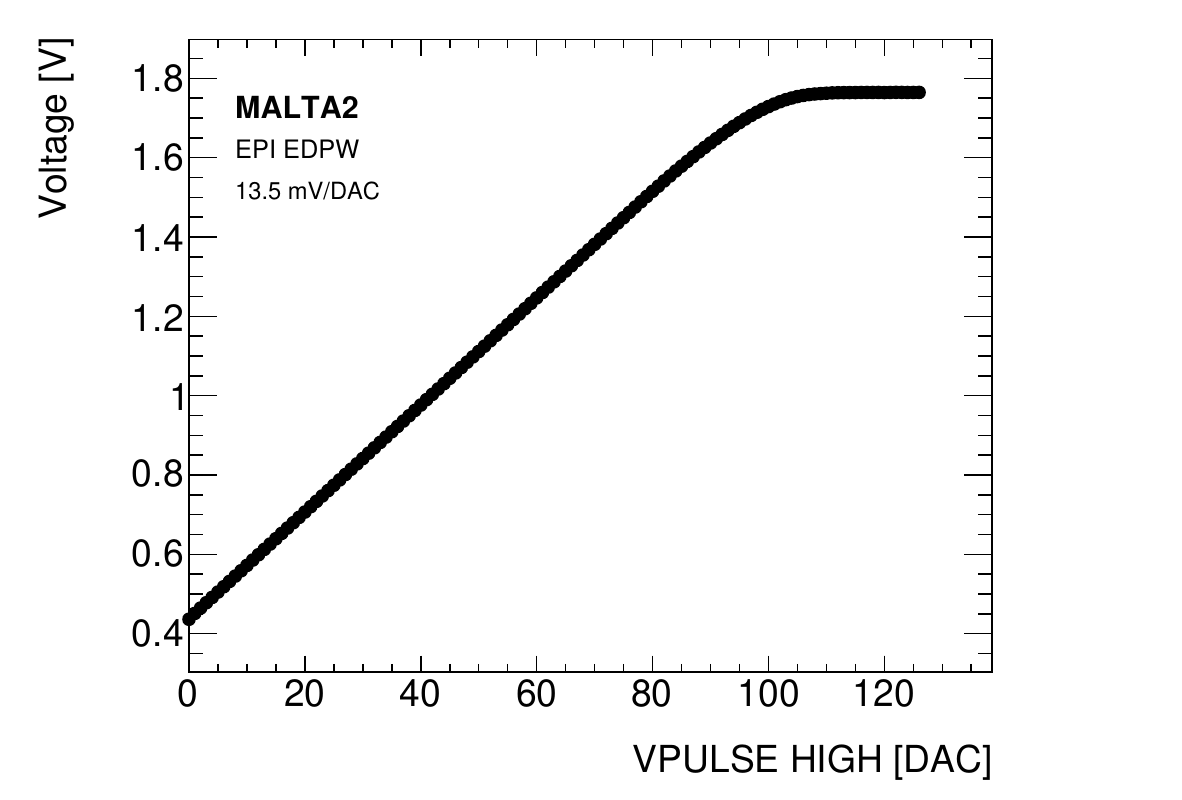}
\caption{$\mathrm{V_{HIGH}}$ voltage characteristic curve generated at the respective DAC value.
The linear responds with a gradient of \SI{13.5}{\milli\volt/DAC} holds up to a DAC value of 90.}
\label{fig:V_HIGH_VDACcurve}
\end{figure}

The voltages $\mathrm{V_{HIGH}}$ and $\mathrm{V_{LOW}}$ are controlled by a DAC with a 7-bit range.
The voltage produced by the DAC as a function of the value of $\mathrm{V_{HIGH}}$ as measured with a Keithley 2400 is shown in figure \ref{fig:V_HIGH_VDACcurve}.
A linear regime is observed from DAC value 0 to 90 with a gradient of \SI{13.5}{\milli\volt/DAC} and an offset of \SI{0.45}{\volt}.
This value will be used to calculate the expected voltage of $\mathrm{V_{HIGH}}$.
The behaviour of the $\mathrm{V_{LOW}}$ DAC is identical.
For DAC values larger than 90 the voltage saturates due to a buffer stage which adds a voltage shift of \SI{0.4}{\volt} and restricts the linear voltage generation. 

\subsection{Digital amplitude of injected charge}
\label{subsec:Digital_Injection}

\begin{figure*}
\centering
\begin{subfigure}[t]{0.49\textwidth}
\centering
\includegraphics[width=\linewidth]{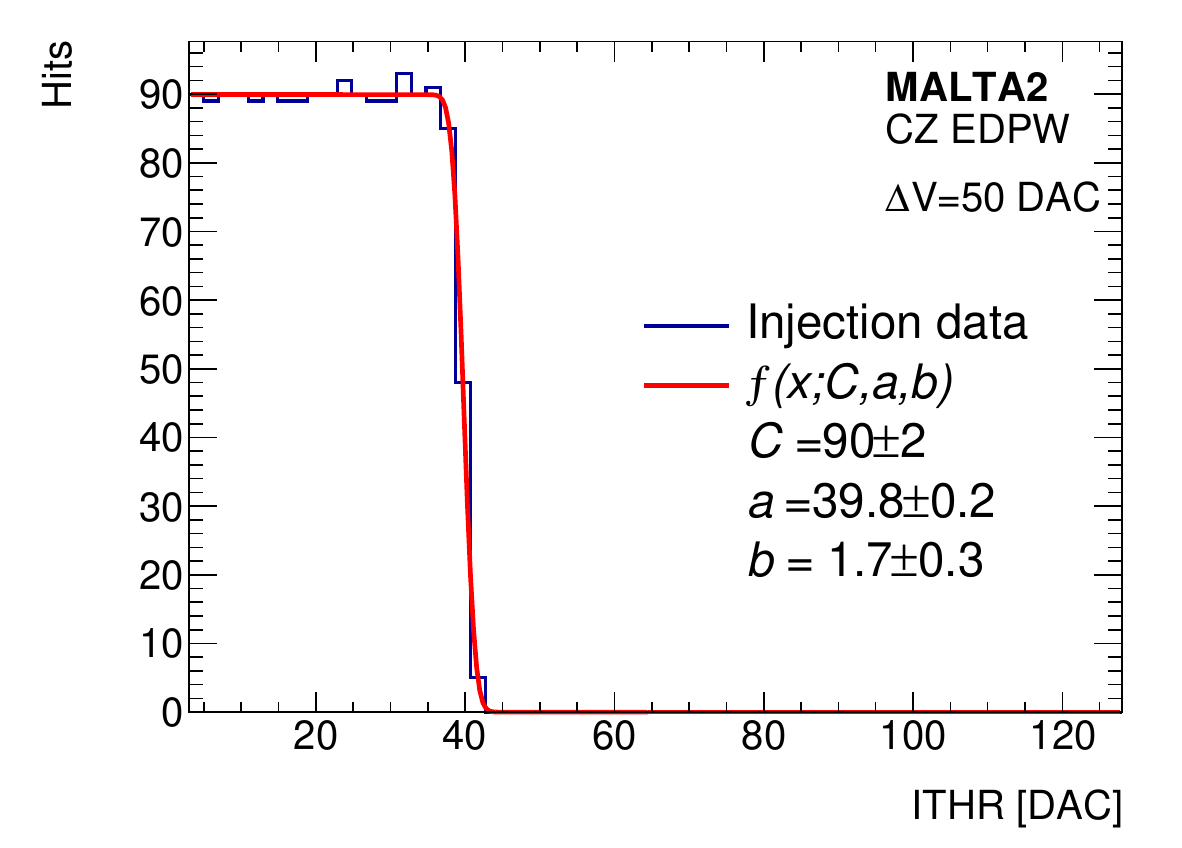}
\caption{Threshold scan of a single pixel under charge injection.}
\label{fig:SinglePix_DigitalAmp}
\end{subfigure}
\begin{subfigure}[t]{0.49\textwidth}
\centering
\includegraphics[width=\linewidth]{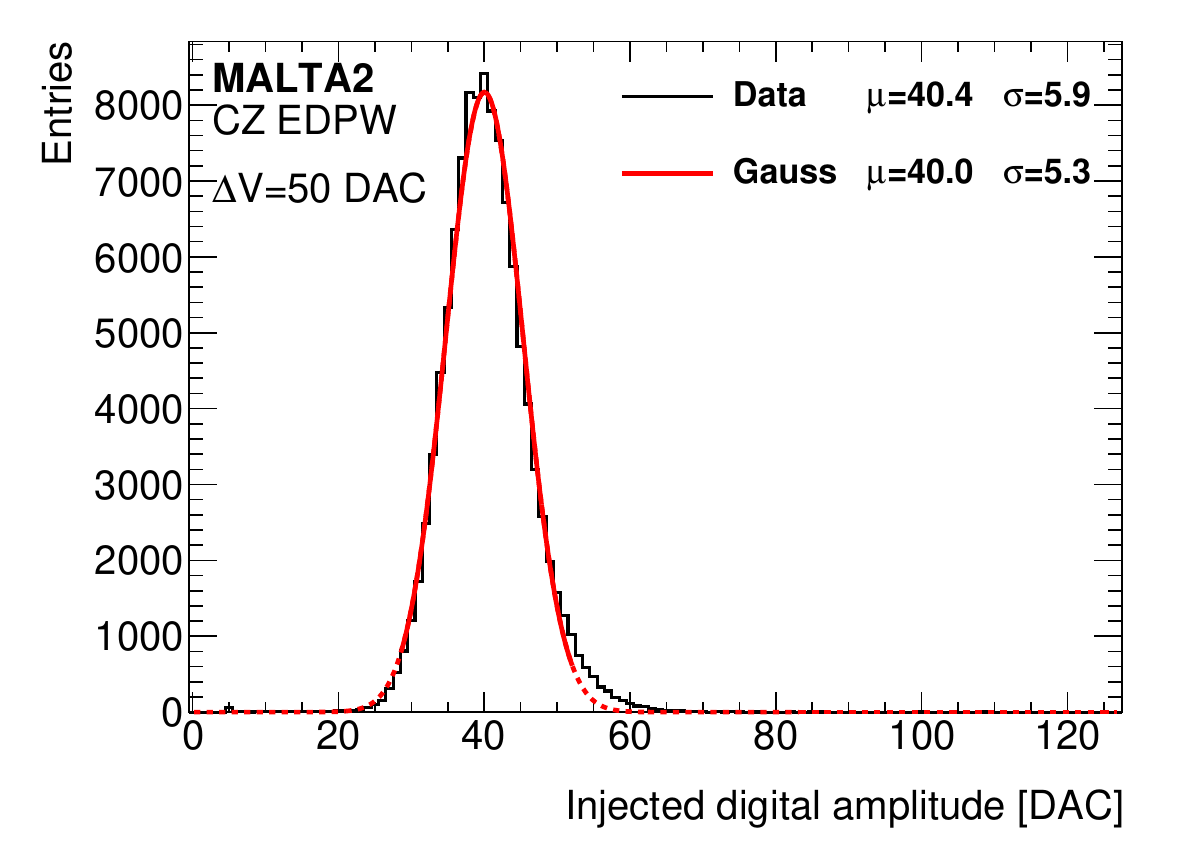}
\caption{Histogram of digital amplitudes for all pixels.}
\label{fig:Histo_digitalamps}
\end{subfigure}
\caption{The digital amplitude from charge injection with $\Delta \mathrm{V}=\mathrm{V_{HIGH}} -\mathrm{V_{LOW}}= 50\,\mathrm{DAC}$ is obtained in (a) for a single pixel through the position parameter $a$ of an error-function fit with width $b$.
The factor $C$ scales with the number of injected pulses.
As the threshold DAC current ITHR is raised above the injected signal the number of detected hits decreases.
The histogram in (b) of digital amplitudes for all $224\times512$ pixels is described by a Gaussian fit to the core of the distribution.
The stated values of the mean $\mu$ and the standard deviation $\sigma$ are those calculated from the data or obtained through the fit.}
\label{fig:DigitalAmpInjection}
\end{figure*}

A threshold scan is a variation of the ITHR current DAC that is proportional to the threshold set in the discriminator in the pixel front-end.
It is proportional to the speed at which the signal returns to the baseline.
From the binary hit data, a signal can be quantified in a threshold scan through the digital amplitude, that is the threshold at which the number of hits are reduced to 50\%.
Figure \ref{fig:SinglePix_DigitalAmp} parameterises the digital amplitude of charge injected into a single pixel through an s-curve 
\begin{equation}
    s(x;C,a,b) = \frac{C}{2}\left[1-\mathrm{erf}\left(\frac{x-a}{\sqrt{2}b}\right)\right]
\end{equation}
with the error-function definition
\begin{equation}
    \mathrm{erf}\left(z\right) = \frac{2}{\sqrt{\pi}}\int_0^z \mathrm{e}^{-t^2} dt.
\end{equation}
Its differentiation 
\begin{equation}
\frac{d}{dx} s(x; C,a,b) = -C \, \mathcal{N}\left(a, b^2\right)
\label{eq:errorfct_diff}
\end{equation}
results in a Gaussian distribution with prefactor $C$, mean $a$ and standard deviation $b$.
The digital amplitude is quantified through the position parameter $a$ of the error-function.
The histogram of the digital amplitudes for all pixels in figure \ref{fig:Histo_digitalamps} is described by a Gaussian distribution with mean amplitude $\mu=40.0$ and standard deviation $\sigma = 5.3$ that quantifies pixel-to-pixel variations.\\

\begin{figure*}
\centering
\begin{subfigure}[t]{0.49\textwidth}
\centering
\includegraphics[width=\linewidth]{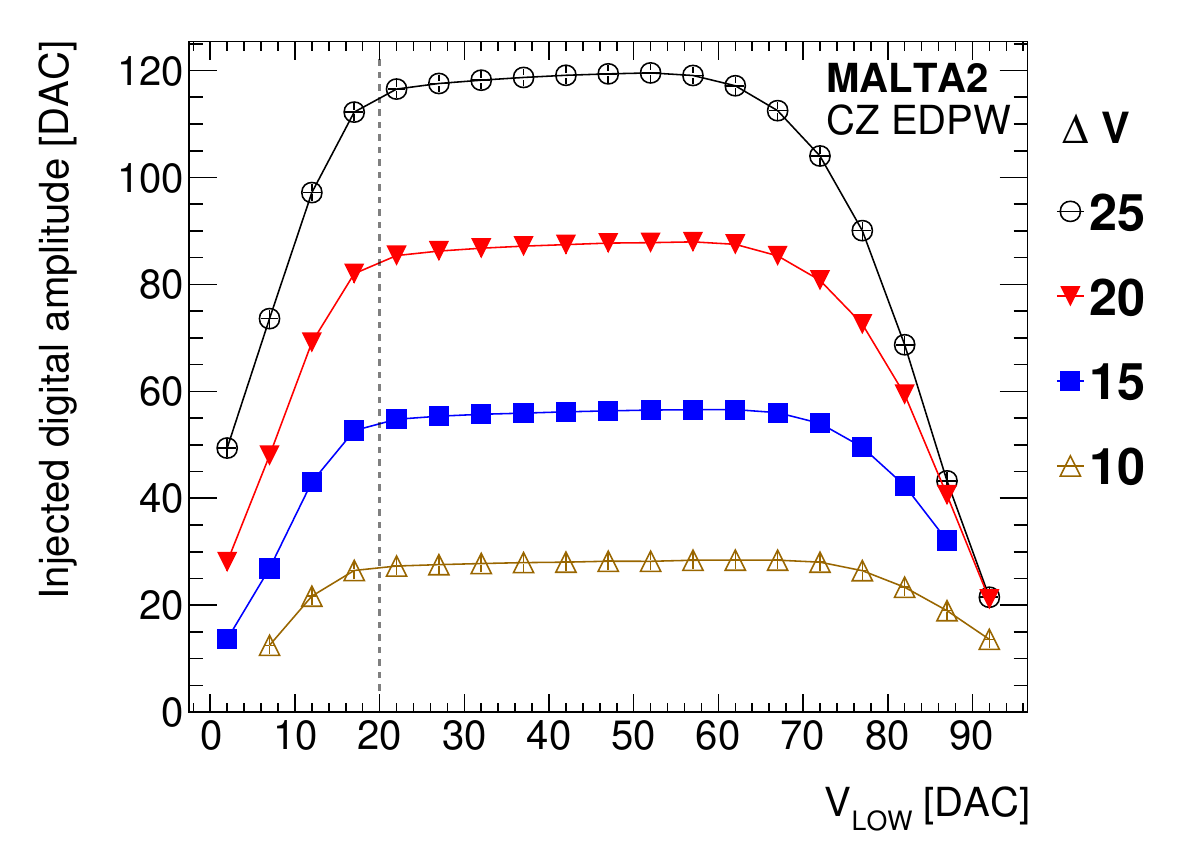}
\caption{Injected digital amplitude versus $\mathrm{V_{LOW}}$.}
\label{fig:VL_validrange}
\end{subfigure}
\begin{subfigure}[t]{0.49\textwidth}
\centering
\includegraphics[width=\linewidth]{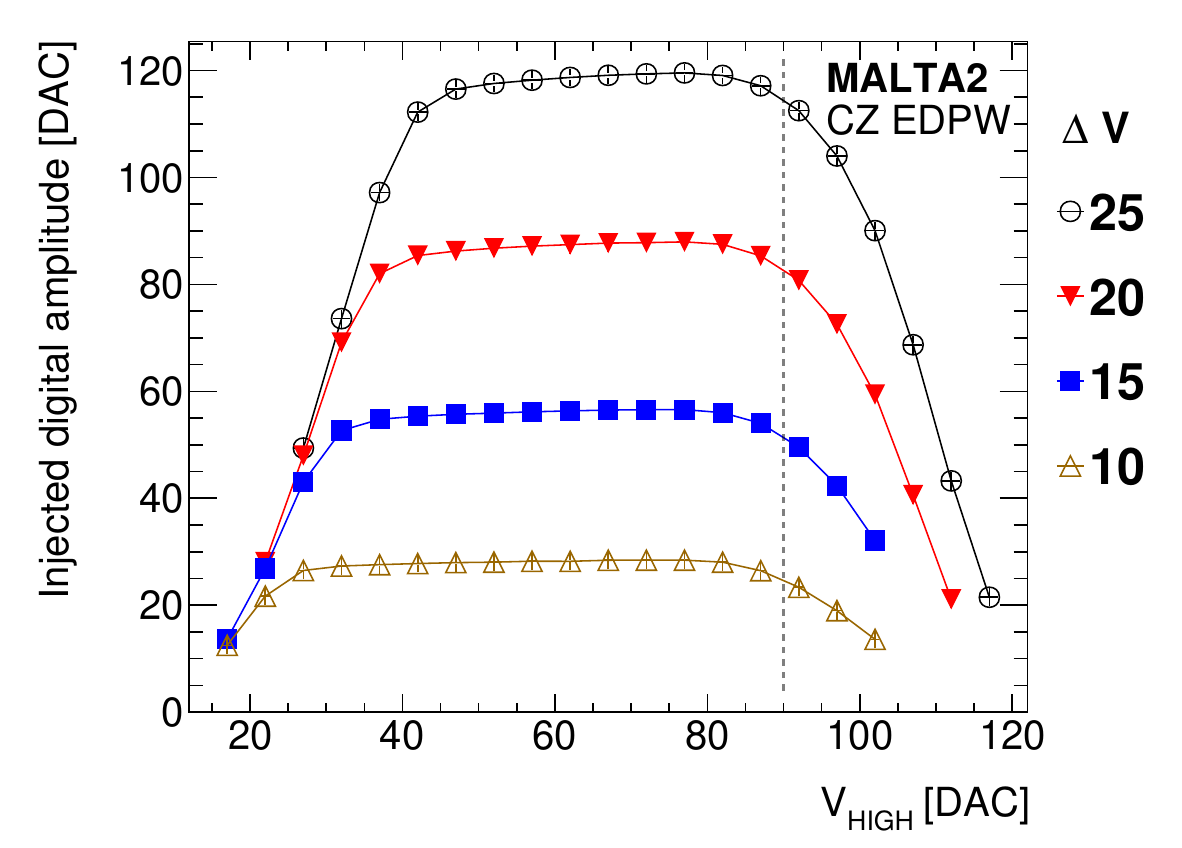}
\caption{Injected digital amplitude versus $\mathrm{V_{HIGH}}$.}
\label{fig:VH_validrange}
\end{subfigure}
\caption{Injected digital amplitude versus $\mathrm{V_{LOW}}$ and $\mathrm{V_{HIGH}}$.
The digital amplitude remains constant in a stable DAC range of [20,90].
The dashed lines mark the lower boundary for $\mathrm{V_{LOW}}$ and the upper boundary for $\mathrm{V_{HIGH}}$.
Outside that range, saturation effects reduce the injected charge which in turn causes a reduction in digital amplitude.}
\label{fig:VH_VL_validrange}
\end{figure*}

According to equation \ref{eq:QInjection}, the injected charge ideally depends on $\Delta \mathrm{V} = \mathrm{V_{HIGH}} -\mathrm{V_{LOW}}$ but not on the individual setting of the two DAC values.
Consequently, a variation of $\mathrm{V_{HIGH}}$ and $\mathrm{V_{LOW}}$ should not affect the digital amplitude when keeping $\Delta \mathrm{V}$ constant.
Figure \ref{fig:VH_VL_validrange} shows the mean digital amplitude as a function of $\mathrm{V_{HIGH}}$ and $\mathrm{V_{LOW}}$ while keeping $\Delta \mathrm{V}$ constant at 10, 15, 20 and 25.
A plateau forms for all $\Delta \mathrm{V}$ which defines the range in which all injections yield the same amplitude. 
At this plateau the amplitude does not depend on the specific value of $\mathrm{V_{HIGH}}$ or $\mathrm{V_{LOW}}$.
The plateau is restricted towards large DAC values by $\mathrm{V_{HIGH}} \leq 90$  and towards small DAC values by $\mathrm{V_{LOW}} \geq 20$.
Both restrictions are indicated by dashed lines.
At low voltages the PMOS transistors that do switching between $\mathrm{V_{HIGH}}$ and $\mathrm{V_{LOW}}$ saturate.
Hence, a reliable injection is obtained when keeping the DAC values within the range [20,90].
From this follows a maximum injection at $\Delta \mathrm{V}=70$.

\subsection{Digital amplitude of Fe-55 source}
\label{subsec:Digital_Fe55source}
\begin{figure}
\centering
\includegraphics[width=\linewidth]{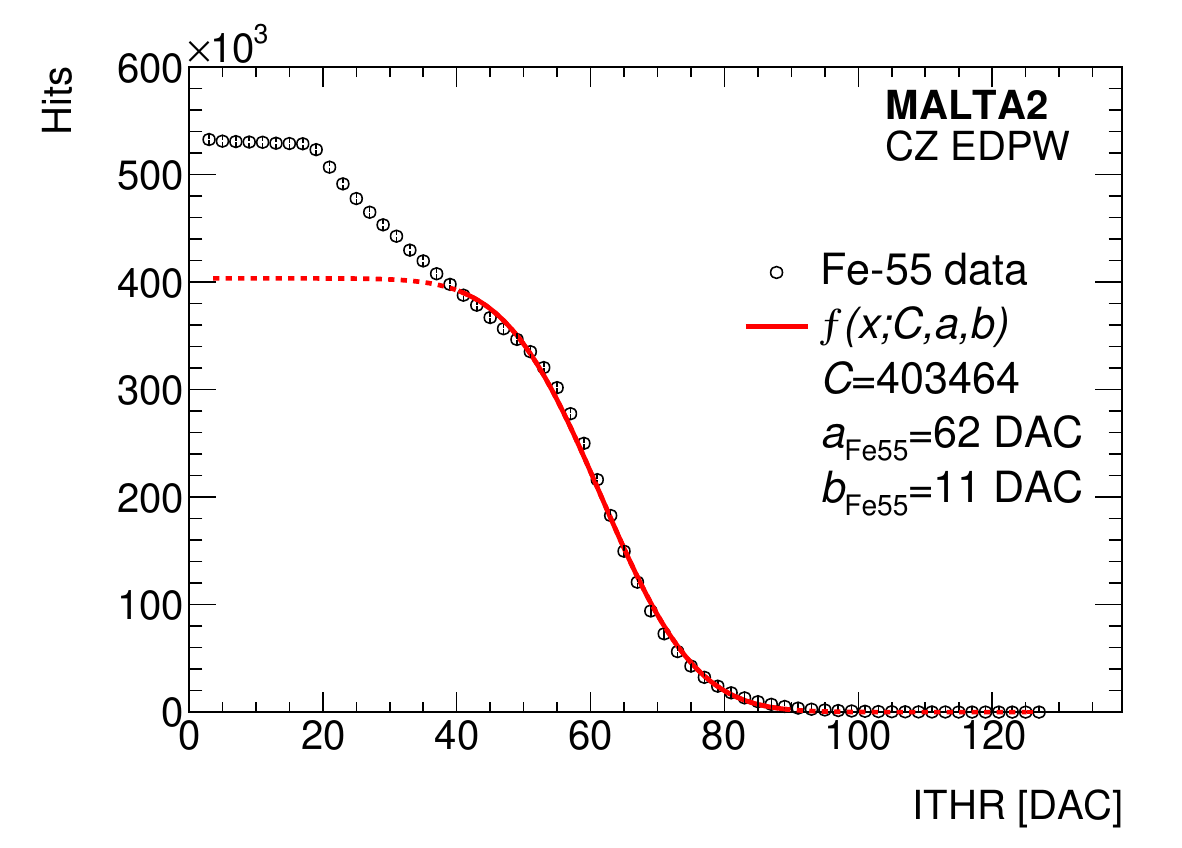}
\caption{Threshold scan under exposure of an Fe-55 source.
Data points (black dots) are the sum of hits from a MALTA2 sensor.
The error-function fit (red line) is fitted to the high threshold regime (solid line) to parameterise the absorption of $\mathrm{K_{\alpha}}$ photons of 5.9 keV. The function is continued
outside the fit regime (dashed line).
The parameter $a_{\mathrm{Fe55}}$ marks the position of the falling edge and $b_{\mathrm{Fe55}}$ the width describing pixel-to-pixel variations.
The deviation of fit function and data at low thresholds is due to charge sharing
where an absorbed photon near a pixel corner causes multiple hits.}
\label{fig:Fe55ITHRScan}
\end{figure}

A gamma source of Fe-55 is used commonly as a calibration source for thin silicon sensors because the charge deposition through photon absorption is around $1600 \,\mathrm{e^-}$ which corresponds to the most probable charge deposition of a minimum-ionizing particle in \SI{27}{\micro\meter} of silicon \cite{Meroli_2011}.
The source emits photons at the $\mathrm{K_{\alpha}}$-lines of \SI{5.9}{\kilo\electronvolt}.
The MALTA2 front-end is designed for thresholds down to $\sim 150\, \mathrm{e^-}$ to ensure high detection efficiency of minimum-ionizing particles \cite{Malta2_PiroFr}.
As a result, the usual threshold range is far below the charge deposition of $1600 \,\mathrm{e^-}$.
However, a low bias current setting in the front-end allows to suppress the gain and configure the threshold to values $>2000\, \mathrm{e^-}$.
In such a low-gain setting, a threshold scan is sensitive to the amplitude deposited by soft X-rays.
A threshold scan of all pixels of a MALTA2 sensor under exposure of an Fe-55 source is shown in figure \ref{fig:Fe55ITHRScan}.
An error-function fit describes the decrease in hits towards large thresholds and yields a digital amplitude of $a_{\mathrm{Fe55}} =62$.
The width parameter $b_{\mathrm{Fe55}} =11$ incorporates pixel-to-pixel variations.

\section{Calibration Results}
\subsection{Charge calibration through Fe-55 source}
\begin{figure}
\centering
\includegraphics[width=\linewidth]{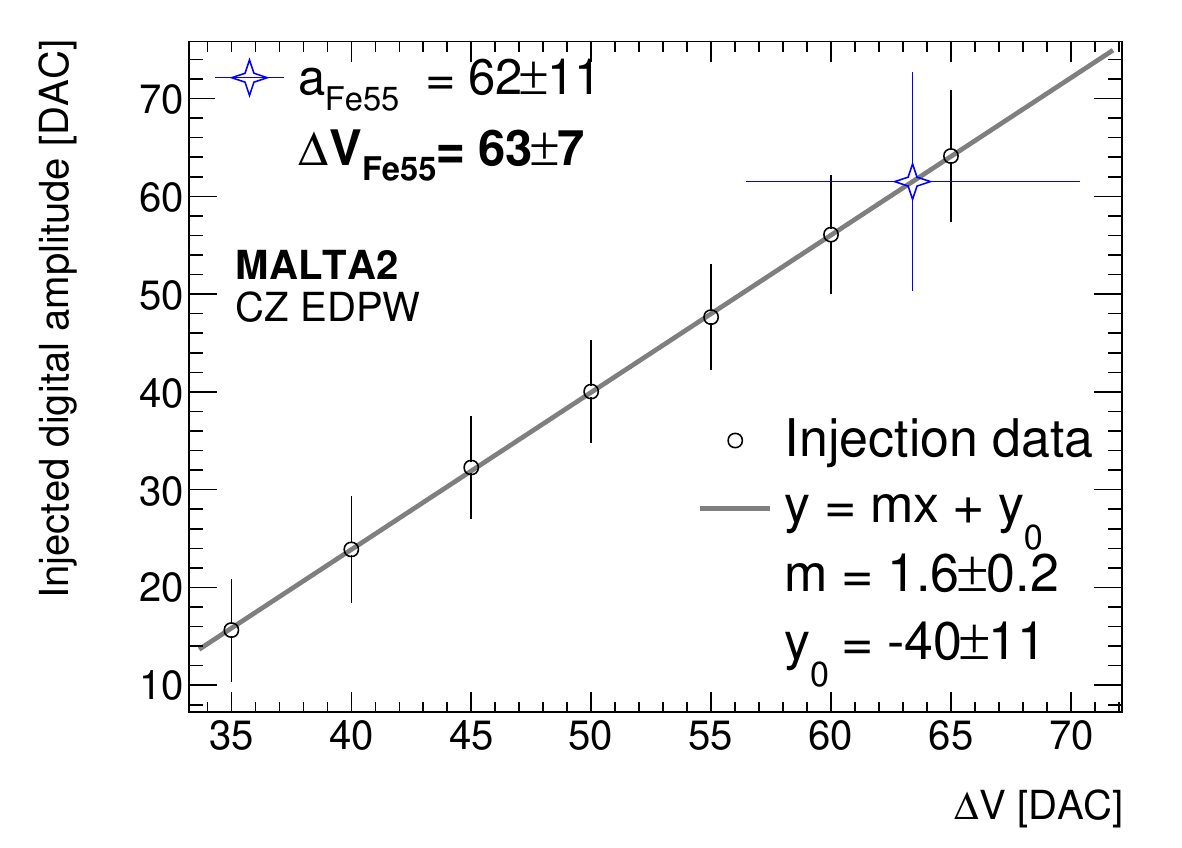}
\caption{Average digital amplitude versus injection DAC setting. 
Data points represent the mean from the Gaussian fit to all pixel amplitudes and the error bars represent the standard deviation. 
The digital amplitude $a_{\mathrm{Fe55}}$ from the Fe-55 detection is added as a calibration point to interpolate from the linear fit the corresponding charge injection $\Delta\mathrm{V_{Fe55}}$ that yields the same digital amplitude.} 
\label{fig:InjectionCalibration}
\end{figure}

Charge is injected for different values of $\Delta \mathrm{V}$ and the mean digital amplitude is reconstructed according to the example for $\Delta \mathrm{V}=50$ in figure \ref{fig:DigitalAmpInjection}.
The amplitude is found to be proportional to $\Delta \mathrm{V}$ as shown in figure \ref{fig:InjectionCalibration}.
The independently measured digital amplitude $a_{\mathrm{Fe55}}$ under the exposure of the Fe-55 source is used as a charge calibration point.
From the linear fit the corresponding $\Delta \mathrm{V_{Fe55}}$ value is calculated. 
The charge injected through the circuit at $\Delta \mathrm{V_{Fe55}}$ yields the same amplitude as $a_{\mathrm{Fe55}}$.

Further, the injection capacitance is calculated as 
\begin{align}
    \mathrm{C_{inj}} &= \frac{1600\, \mathrm{e^-} \times \SI{1.602E-19}{\coulomb\per \mathrm{e^-}}}{\Delta \mathrm{V_{Fe55}} \times \SI{13.5}{\milli\volt/DAC}} \nonumber \\
    &= \frac{\SI{19000}{\atto\farad}}{\Delta \mathrm{V_{Fe55}/DAC}}
\end{align}
by assuming a voltage input per unit $\Delta \mathrm{V}$ in the injection circuit of \SI{13.5}{\milli\volt/DAC}.

Table \ref{tab:CalibResults} lists the calibration results of $\Delta \mathrm{V_{Fe55}}$ and $\mathrm{C_{inj}}$ for all tested samples.
Considering only the six non irradiated samples an average injection capacitance of
\begin{equation}
    \mathrm{C_{inj,exp}}=\SI{257}{\atto\farad}
\end{equation}
is found to be 12\% larger than the design value of $\mathrm{C_{inj}}=\SI{230}{\atto\farad}$.
For the Fe-55 equivalent injection DAC the average value is determined to be
\begin{equation}
    \Delta \mathrm{V_{Fe55,exp}} = 75 \pm 10 \,\mathrm{DAC}.
\end{equation}
The stated uncertainty is the standard deviation of the sample and estimates the fluctuation among different sensors originating from the fabrication process.

\begin{table*}
 \centering 
 \caption{MALTA2 Calibration Results}
\begin{tabular}{ccScc}
    \toprule
    sample & doping & {NIEL} & {$\Delta \mathrm{V_{Fe55}}$}& {$\mathrm{C_{inj}}$}\\
    & &  {[$10^{15}\,\mathrm{n_{eq}}\si{\per \square\centi\meter}$]} & {[DAC]} & {[aF]}\\
    \midrule
    W5R21 & high & 0& $76 \pm 7$& $250 \pm 23$ \\
    W8R24 & low  & 0& $87 \pm 9$& $218 \pm 23$ \\
    W11R0 & high & 0& $63 \pm 7$& $301 \pm 33$ \\
    W14R11& high & 0& $85 \pm 8$& $223 \pm 21$ \\
    W18R17 & very high & 0& $75\pm 6$& $253 \pm 20$ \\
    W18R19 & very high & 0& $64\pm 5$& $297 \pm 23$ \\
    \hline
    W12R7 & high & 1& $66\pm 8$& $288 \pm 35$ \\
    W18R1 & very high & 1& $65\pm 8$& $ 292\pm 36$ \\
    W18R4 & very high & 2& $58\pm 8$& $327 \pm 45$ \\
    W18R9 & very high & 3& $47\pm 6$& $ 404\pm 52$ \\
    W18R21 & very high & 3& $53\pm 6$& $358 \pm 41$ \\
    W18R12 & very high & 5& $46\pm 10$& $413 \pm 90$ \\
    W18R14 & very high & 5& $41\pm 8$& $463 \pm 90$ \\
    \botrule
   \end{tabular}
   \label{tab:CalibResults}
 \end{table*}

\subsection{Calibrated threshold scans}
Based on the result of $\Delta \mathrm{V_{Fe55}}$ any charge injected into a MALTA2 sensor can be calibrated to unit electrons according to 
\begin{equation}
    \mathrm{Q_{inj}} = \Delta \mathrm{V}\, \frac{1600\, \mathrm{e^-}}{\Delta \mathrm{V_{Fe55}}}.
    \label{eq:calibratedQ}
\end{equation}

\begin{figure*}
\centering
\begin{subfigure}[t]{0.49\textwidth}
\centering
\includegraphics[width=\linewidth]
{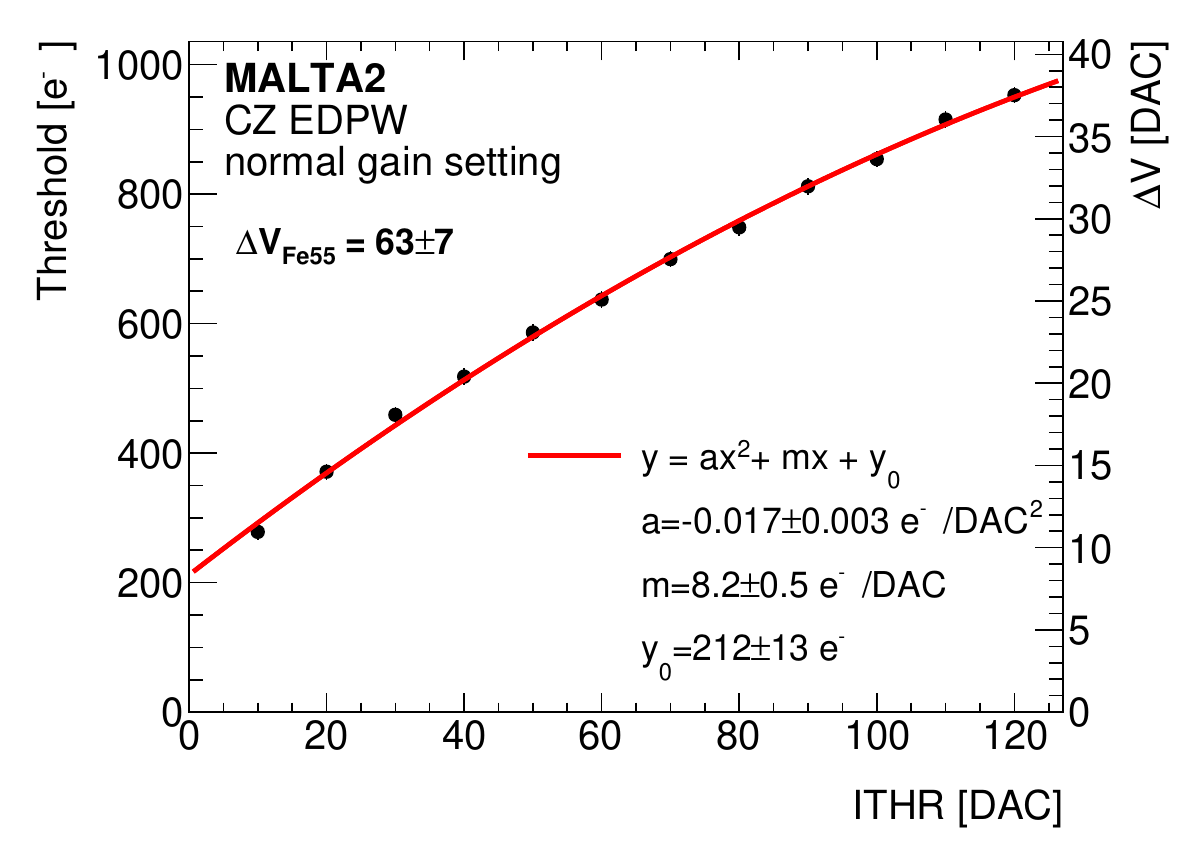}
\caption{Thresholds for front-end at normal gain.}
\label{fig:ThreshScan_normalgain}
\end{subfigure}
\begin{subfigure}[t]{0.49\textwidth}
\centering
\includegraphics[width=\linewidth]
{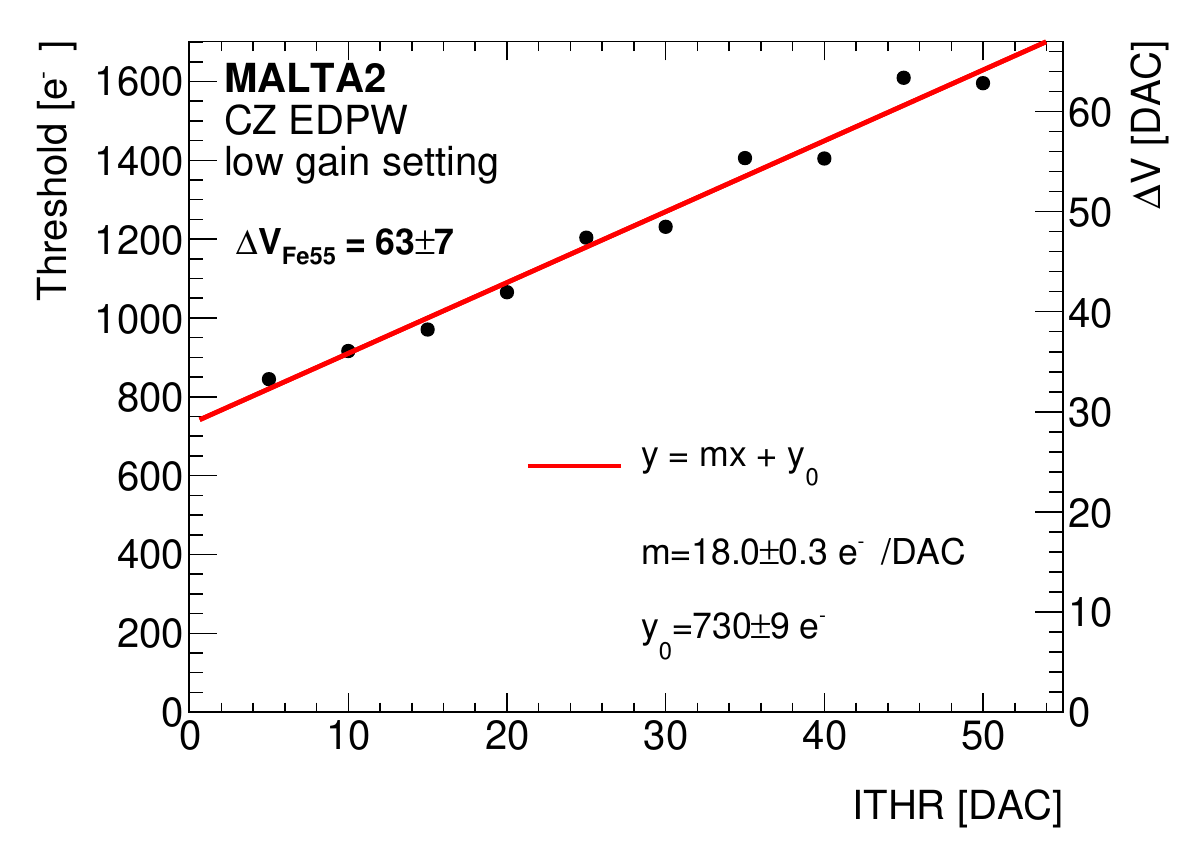}
\caption{Thresholds for front-end at low gain.}
\label{fig:ThreshScan_lowgain}
\end{subfigure}
\caption{Calibrated threshold as a function of the ITHR DAC.
For each ITHR value, the threshold is measured per pixel in units of injected charge $\Delta V$ (right y-axes). 
The calibration input of $\Delta \mathrm{V_{Fe55}}$ makes a linear calibration to unit electrons according to equation \ref{eq:calibratedQ} possible (left y-axes).
(a) shows a normal gain setting at IBIAS=43 with thresholds down to 200 $\mathrm{e^-}$ and a parabolic fit.
(b) shows a  low gain front-end setting at IBIAS=3 with thresholds above 750 $\mathrm{e^-}$ and a linear fit.}
\label{fig:ThreshScans}
\end{figure*}

For a normal and low gain front-end setting the detection threshold is measured as an average over all pixels and is shown in figure \ref{fig:ThreshScans}.
The low gain is the result of reducing the main front-end bias current IBIAS by a factor of 14 compared to the normal setting and leads to larger threshold values.
The thresholds in unit electrons (left y-axes) are obtained through equation \ref{eq:calibratedQ} based on the measurement of $\Delta V$ (right y-axes).
Error bars quantify the error on the mean threshold and are of the same order than the marker size.
The threshold dependence on ITHR can be parameterised by a linear or quadratic function.
These threshold values define the detection threshold at testbeam studies and beam telescope applications \cite{vanRijnbach:2023gtl, Gustavino:2024qj}.
The threshold resolution quantifying the standard deviation from pixel to pixel variations is around 10\%. 

\subsection{Irradiation study}
\begin{figure*}
\centering
\begin{subfigure}[t]{0.49\textwidth}
\centering
\includegraphics[width=\linewidth]{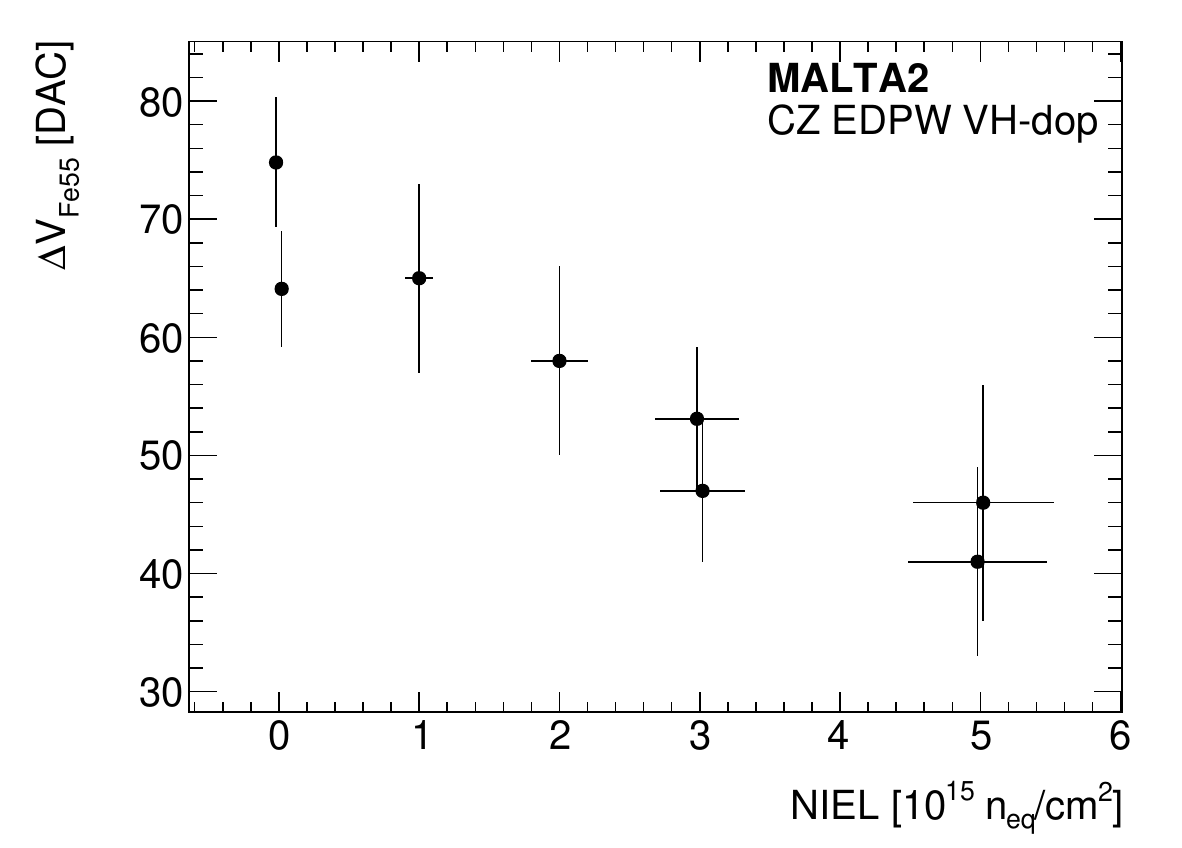}
\caption{$\Delta \mathrm{V_{Fe55}}$ versus NIEL.}
\label{fig:Irrad_Comparison_dVFe55}
\end{subfigure}
\begin{subfigure}[t]{0.49\textwidth}
\centering
\includegraphics[width=\linewidth]{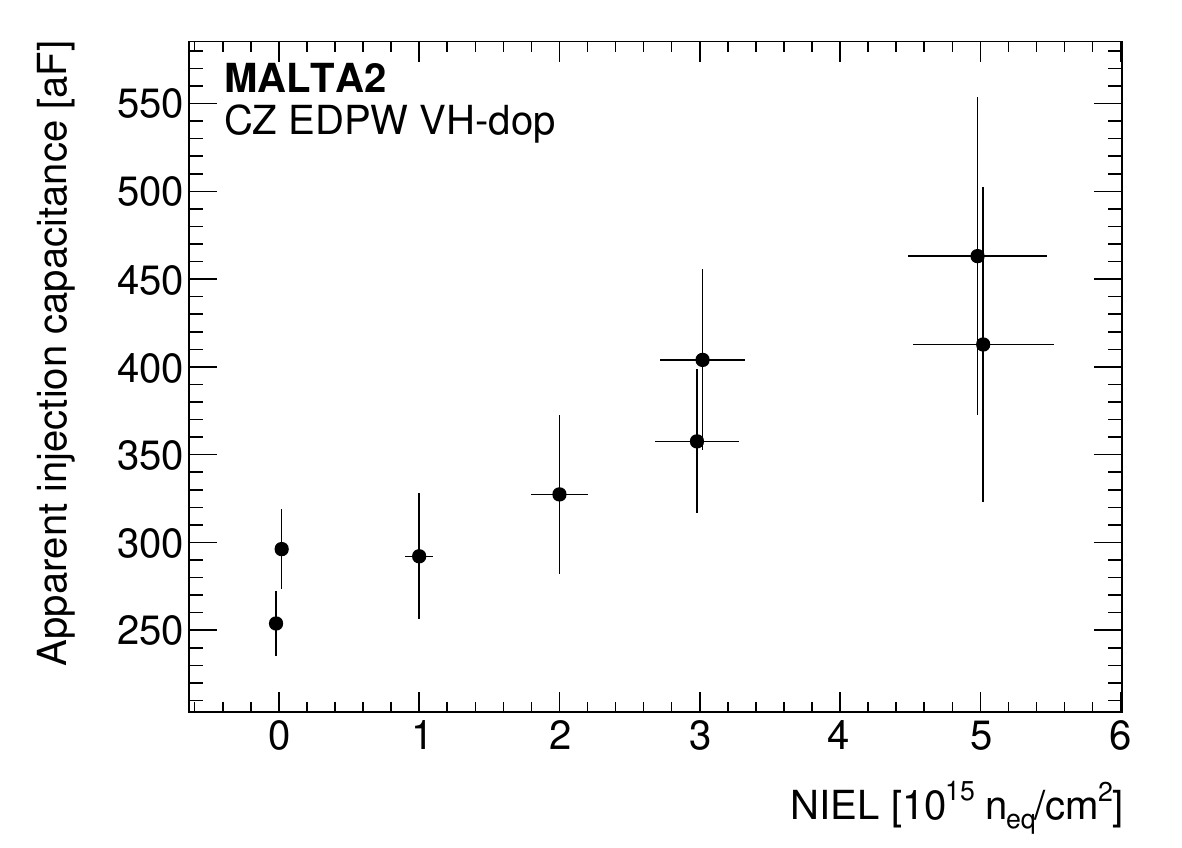}
\caption{Apparent injection capacitance versus NIEL.}
\label{fig:Irrad_Comparison_InjCap}
\end{subfigure}
\caption{Dependence of $\Delta \mathrm{V_{Fe55}}$ (a) and the corresponding injection capacitance (b) on the irradiation fluence as non-ionizing energy loss (NIEL). All samples originate from the same wafer of \SI{100}{\micro\meter} thick Czochralski silicon with very high doping of n- layer.
$\Delta \mathrm{V_{Fe55}}$ is the voltage DAC value that injects a charge of $1600\, \mathrm{e^-}$ corresponding to the main Fe-55 deposition.
$\Delta \mathrm{V_{Fe55}}$ decreases with irradiation.
Consequently, the apparent injection capacitance increases with irradiation because a lower voltage is needed to inject the same reference charge.
}
\label{fig:Irrad_Comparison_W18}
\end{figure*}

A selection of samples from the same wafer of \SI{100}{\micro\meter} thick Czochralski silicon with very high doping of the n- layer has been neutron irradiated at different fluences of non-ionizing energy loss (NIEL) at the Triga reactor in the Institute Jožef Stefan (IJS), Slovenia.
The calibration results are compared in figure \ref{fig:Irrad_Comparison_W18} and show a decrease in the Fe-55 equivalent injection voltage $\Delta \mathrm{V_{Fe55}}$ with fluence.
In irradiated samples, a lower injection voltage is needed to inject  1600 $\mathrm{e^-}$.
This can be interpreted as an increase in the apparent injection capacitance. 
Two sensors are tested for each of the fluences at 1, 3 and $5\times10^{15}\,\mathrm{n_{eq}}\si{\per\square\centi\meter}$ and show compatibility within the order of the stated uncertainties.
The study shows that charge calibration has to account for the irradiation fluence for precise threshold determination.

\section{Summary} 
\label{sec:summary}
The charge injection circuit of MALTA2 sensors has been calibrated inside a reliable DAC range of [20,90] for the parameters $\mathrm{V_{HIGH}}$ and $\mathrm{V_{LOW}}$.
A digital amplitude is reconstructed through threshold scans from binary hit data as the position parameter of an error-function fit.
The amplitude is proportional to the voltage input of the charge injection circuit 
$\Delta \mathrm{V} = \mathrm{V_{HIGH}} - \mathrm{V_{LOW}}$ and is calibrated through an Fe-55 source.
Based on this, the charge injected into a MALTA2 sensor can be calculated:
\begin{equation}
    \mathrm{Q_{inj}} = \Delta \mathrm{V}\, \frac{1600\, \mathrm{e^-}}{\Delta \mathrm{V_{Fe55}}}
    \label{eq:calibratedQ2}
\end{equation}
It is proposed that for precise calibration $\Delta \mathrm{V_{Fe55}}$ is determined for each sample. 
For the six tested non irradiated sensors the mean and standard deviation of $\Delta \mathrm{V_{Fe55,exp}} = 75 \pm 10$ is obtained.
The mean value is assumed for calibration of non irradiated samples that are not calibrated individually.
Neutron irradiated samples show a smaller value of $\mathrm{V_{Fe55}}$ due to an increase in the effective injection capacitance.
Once the charge calibration parameter $\Delta \mathrm{V_{Fe55}}$ is obtained, the threshold setting of a MALTA2 sensor is quantifiable at any front-end setting through charge injection.
The described procedure is feasible for any binary sensor that incorporates an injection circuit.

\backmatter

\bmhead{Acknowledgements}
This project has received funding from the European Union’s Horizon 2020 Research and Innovation programme under Grant Agreement number 101004761 (AIDAinnova), and number 654168 (IJS, Ljubljana, Slovenia). Furthermore it has been supported by the Marie Sklodowska-Curie Innovative Training Network of the European Commission Horizon 2020 Programme under contract number 675587 (STREAM).
This publication was funded by Deutsche Forschungsgemeinschaft (DFG, German Research Foundation) - 491245950.

\bibliography{malta}


\begin{thebibliography}{10}
\ifx \bisbn   \undefined \def \bisbn  #1{ISBN #1}\fi
\ifx \binits  \undefined \def \binits#1{#1}\fi
\ifx \bauthor  \undefined \def \bauthor#1{#1}\fi
\ifx \batitle  \undefined \def \batitle#1{#1}\fi
\ifx \bjtitle  \undefined \def \bjtitle#1{#1}\fi
\ifx \bvolume  \undefined \def \bvolume#1{\textbf{#1}}\fi
\ifx \byear  \undefined \def \byear#1{#1}\fi
\ifx \bissue  \undefined \def \bissue#1{#1}\fi
\ifx \bfpage  \undefined \def \bfpage#1{#1}\fi
\ifx \blpage  \undefined \def \blpage #1{#1}\fi
\ifx \burl  \undefined \def \burl#1{\textsf{#1}}\fi
\ifx \doiurl  \undefined \def \doiurl#1{\url{https://doi.org/#1}}\fi
\ifx \betal  \undefined \def \betal{\textit{et al.}}\fi
\ifx \binstitute  \undefined \def \binstitute#1{#1}\fi
\ifx \binstitutionaled  \undefined \def \binstitutionaled#1{#1}\fi
\ifx \bctitle  \undefined \def \bctitle#1{#1}\fi
\ifx \beditor  \undefined \def \beditor#1{#1}\fi
\ifx \bpublisher  \undefined \def \bpublisher#1{#1}\fi
\ifx \bbtitle  \undefined \def \bbtitle#1{#1}\fi
\ifx \bedition  \undefined \def \bedition#1{#1}\fi
\ifx \bseriesno  \undefined \def \bseriesno#1{#1}\fi
\ifx \blocation  \undefined \def \blocation#1{#1}\fi
\ifx \bsertitle  \undefined \def \bsertitle#1{#1}\fi
\ifx \bsnm \undefined \def \bsnm#1{#1}\fi
\ifx \bsuffix \undefined \def \bsuffix#1{#1}\fi
\ifx \bparticle \undefined \def \bparticle#1{#1}\fi
\ifx \barticle \undefined \def \barticle#1{#1}\fi
\bibcommenthead
\ifx \bconfdate \undefined \def \bconfdate #1{#1}\fi
\ifx \botherref \undefined \def \botherref #1{#1}\fi
\ifx \url \undefined \def \url#1{\textsf{#1}}\fi
\ifx \bchapter \undefined \def \bchapter#1{#1}\fi
\ifx \bbook \undefined \def \bbook#1{#1}\fi
\ifx \bcomment \undefined \def \bcomment#1{#1}\fi
\ifx \oauthor \undefined \def \oauthor#1{#1}\fi
\ifx \citeauthoryear \undefined \def \citeauthoryear#1{#1}\fi
\ifx \endbibitem  \undefined \def \endbibitem {}\fi
\ifx \bconflocation  \undefined \def \bconflocation#1{#1}\fi
\ifx \arxivurl  \undefined \def \arxivurl#1{\textsf{#1}}\fi
\csname PreBibitemsHook\endcsname

\bibitem[\protect\citeauthoryear{Pernegger et~al.}{2017}]{Pernegger_2017}
\begin{barticle}
\bauthor{\bsnm{Pernegger}, \binits{H.}}, \betal:
\batitle{First tests of a novel radiation hard {CMOS} sensor process for
  {Depleted Monolithic Active Pixel Sensors}}.
\bjtitle{JINST}
\bvolume{12}(\bissue{06}),
\bfpage{06008}
(\byear{2017})
\doiurl{10.1088/1748-0221/12/06/P06008}
\end{barticle}
\endbibitem

\bibitem[\protect\citeauthoryear{Berdalovic et~al.}{2018}]{Berdalovic:2018tce}
\begin{bchapter}
\bauthor{\bsnm{Berdalovic}, \binits{I.}}, \betal:
\bctitle{{MALTA: a CMOS pixel sensor with asynchronous readout for the ATLAS
  High-Luminosity upgrade}}.
In: \bbtitle{Proc. IEEE Nucl. Sci. Symp. Med. Imag. Conf.},
p. \bfpage{8824349}
(\byear{2018}).
\doiurl{10.1109/NSSMIC.2018.8824349}
\end{bchapter}
\endbibitem

\bibitem[\protect\citeauthoryear{Piro et~al.}{2022}]{Malta2_PiroFr}
\begin{botherref}
\oauthor{\bsnm{Piro}, \binits{F.}}, et al.:
A 1 {$\mathrm{\mu}$W} radiation-hard front-end in a 0.18 {$\mathrm{\mu m}$}
  {CMOS} process for the {MALTA2} monolithic sensor.
IEEE Trans. Nucl. Sci.
\textbf{69}(6)
(2022)
\doiurl{10.1109/TNS.2022.3170729}
\end{botherref}
\endbibitem

\bibitem[\protect\citeauthoryear{{van Rijnbach}
  et~al.}{2024}]{vanRijnbach:2023gtl}
\begin{barticle}
\bauthor{\bsnm{{van Rijnbach}}, \binits{M.}}, \betal:
\batitle{{Radiation hardness of MALTA2 monolithic CMOS imaging sensors on
  Czochralski substrates}}.
\bjtitle{Eur. Phys. J. C}
\bvolume{84}(\bissue{3}),
\bfpage{251}
(\byear{2024})
\doiurl{10.1140/epjc/s10052-024-12601-3}
{\href{https://arxiv.org/abs/2308.13231}{{arXiv:2308.13231}}}
\end{barticle}
\endbibitem

\bibitem[\protect\citeauthoryear{Berdalovic}{2019}]{Berdalovic:2702884}
\begin{botherref}
\oauthor{\bsnm{Berdalovic}, \binits{I.}}:
{Design of radiation-hard CMOS sensors for particle detection applications}.
PhD thesis,
Zagreb U.
(2019).
Presented 29 Oct 2019.
\url{http://cds.cern.ch/record/2702884}
\end{botherref}
\endbibitem

\bibitem[\protect\citeauthoryear{Pohl et~al.}{2015}]{ThresholdMethod_Reconstr}
\begin{barticle}
\bauthor{\bsnm{Pohl}, \binits{D.-L.}}, \betal:
\batitle{Obtaining spectroscopic information with the {ATLAS FE-I4} pixel
  readout chip}.
\bjtitle{NIMA}
\bvolume{788},
\bfpage{49}--\blpage{53}
(\byear{2015})
\doiurl{10.1016/j.nima.2015.03.067}
\end{barticle}
\endbibitem

\bibitem[\protect\citeauthoryear{Fasselt
  et~al.}{2023}]{10.3389/fphy.2023.1231336}
\begin{botherref}
\oauthor{\bsnm{Fasselt}, \binits{L.}}, et al.:
Energy calibration through {X-ray} absorption of the {DECAL} sensor, a
  monolithic active pixel sensor prototype for digital electromagnetic
  calorimetry and tracking.
Frontiers in Physics
\textbf{11}
(2023)
\doiurl{10.3389/fphy.2023.1231336}
\end{botherref}
\endbibitem

\bibitem[\protect\citeauthoryear{McCormack
  et~al.}{2020}]{mccormack2020newmethodsiliconsensor}
\begin{botherref}
\oauthor{\bsnm{McCormack}, \binits{P.}}, et al.:
New Method for Silicon Sensor Charge Calibration Using Compton Scattering
(2020).
\url{https://arxiv.org/abs/2008.11860}
\end{botherref}
\endbibitem

\bibitem[\protect\citeauthoryear{Meroli et~al.}{2011}]{Meroli_2011}
\begin{barticle}
\bauthor{\bsnm{Meroli}, \binits{S.}}, \betal:
\batitle{Energy loss measurement for charged particles in very thin silicon
  layers}.
\bjtitle{JINST}
\bvolume{6}(\bissue{06}),
\bfpage{06013}
(\byear{2011})
\doiurl{10.1088/1748-0221/6/06/P06013}
\end{barticle}
\endbibitem

\bibitem[\protect\citeauthoryear{Gustavino et~al.}{2024}]{Gustavino:2024qj}
\begin{bchapter}
\bauthor{\bsnm{Gustavino}, \binits{G.}}, \betal:
\bctitle{{Development of the radiation-hard MALTA CMOS sensor for tracking
  applications}}.
In: \bbtitle{Proc. Vertex 2023},
vol. \bseriesno{448},
p. \bfpage{048}
(\byear{2024}).
\doiurl{10.22323/1.448.0048}
\end{bchapter}
\endbibitem

\end{thebibliography}


\begin{appendices}
\section{Calibrated samples}\label{secA1}
\begin{figure*}
\centering
\begin{subfigure}[t]{0.49\textwidth}
\centering
\includegraphics[width=\linewidth]{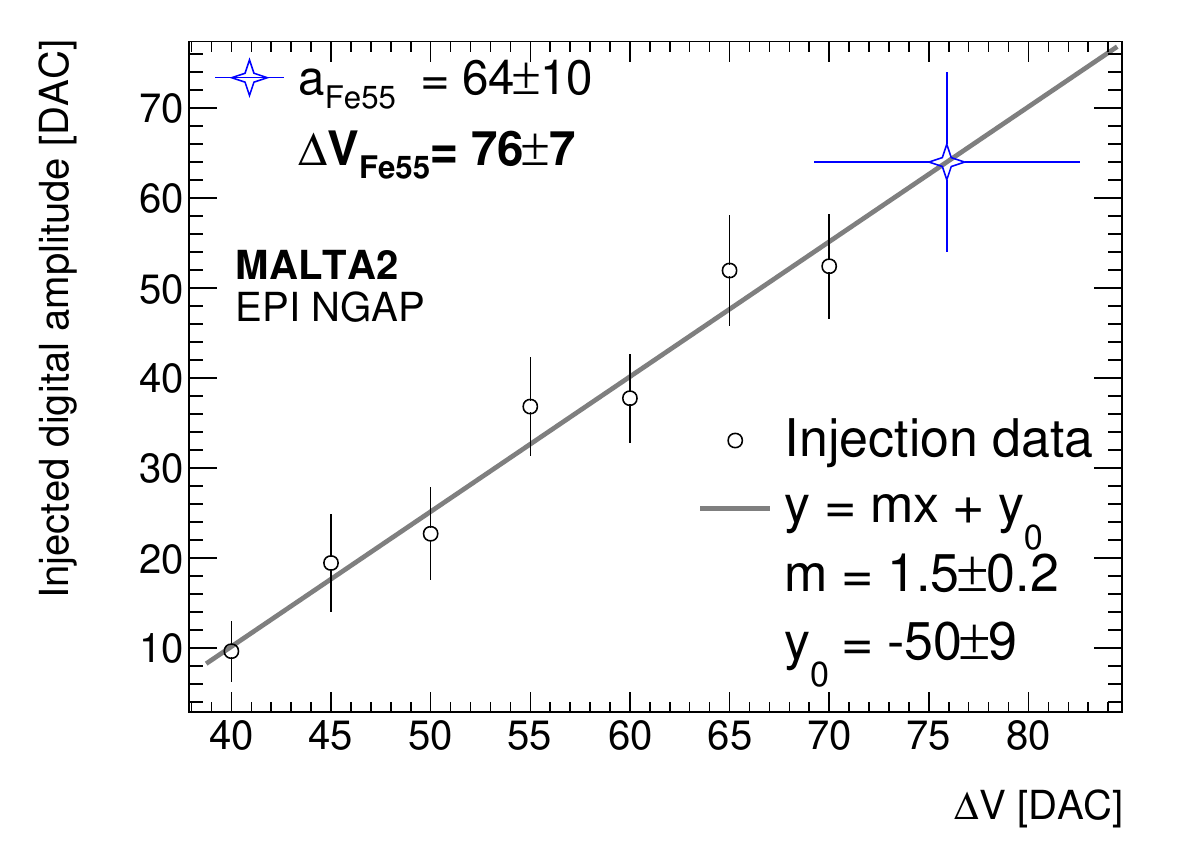}
\caption{W5R21}
\end{subfigure}
\begin{subfigure}[t]{0.49\textwidth}
\centering
\includegraphics[width=\linewidth]{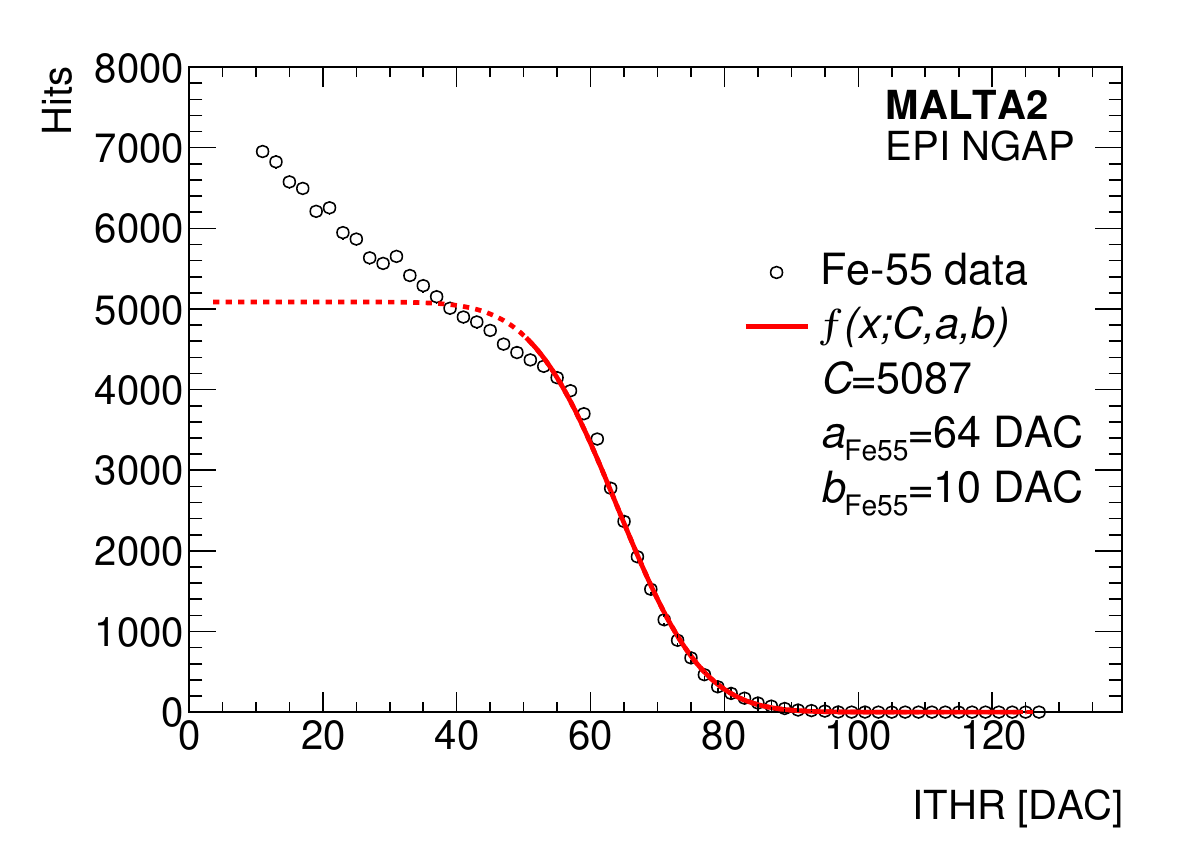}
\caption{W5R21}
\end{subfigure}

\begin{subfigure}[t]{0.49\textwidth}
\centering
\includegraphics[width=\linewidth]{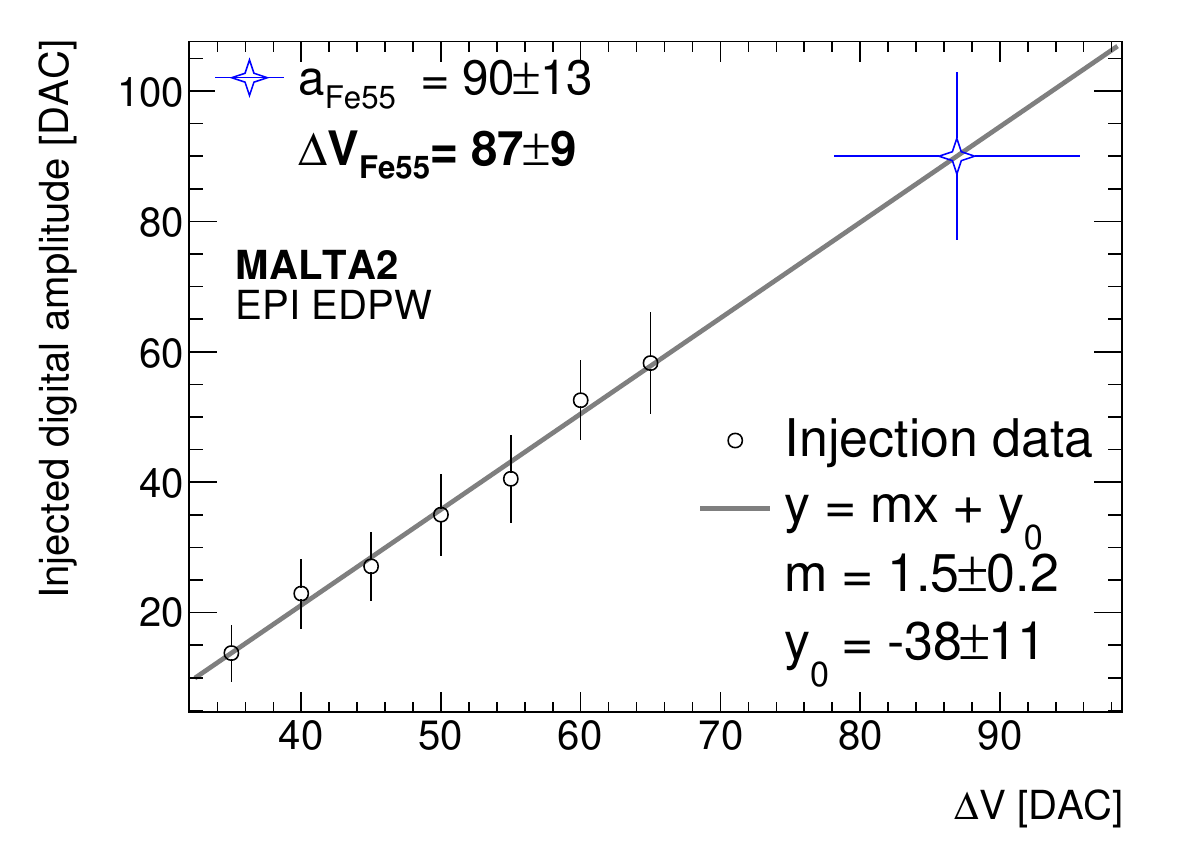}
\caption{W8R24}
\end{subfigure}
\begin{subfigure}[t]{0.49\textwidth}
\centering
\includegraphics[width=\linewidth]{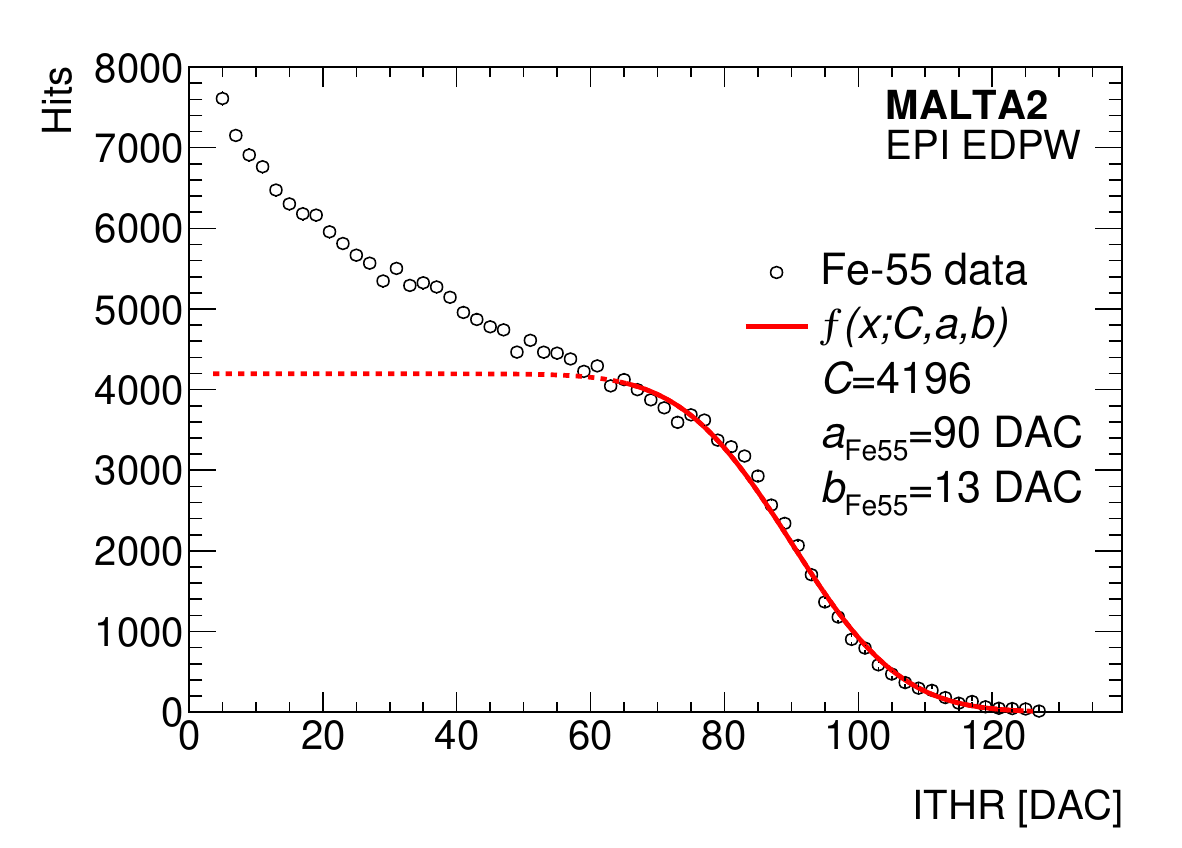}
\caption{W8R24}
\end{subfigure}

\begin{subfigure}[t]{0.49\textwidth}
\centering
\includegraphics[width=\linewidth]{Appendix_plots/Inj_W11R0_MeanDigSignals_DACmin20_DACmax90__IBIAS03_IDB100_ITHR_LinFit.pdf}
\caption{W11R0}
\end{subfigure}
\begin{subfigure}[t]{0.49\textwidth}
\centering
\includegraphics[width=\linewidth]{Appendix_plots/Fe55Hits_ErfFit_W11R0_Fe55source_N10_n100000.pdf}
\caption{W11R0}
\end{subfigure}

\caption{}
\label{fig:Append_Calib1}
\end{figure*}

\begin{figure*}
\centering
\begin{subfigure}[t]{0.49\textwidth}
\centering
\includegraphics[width=\linewidth]{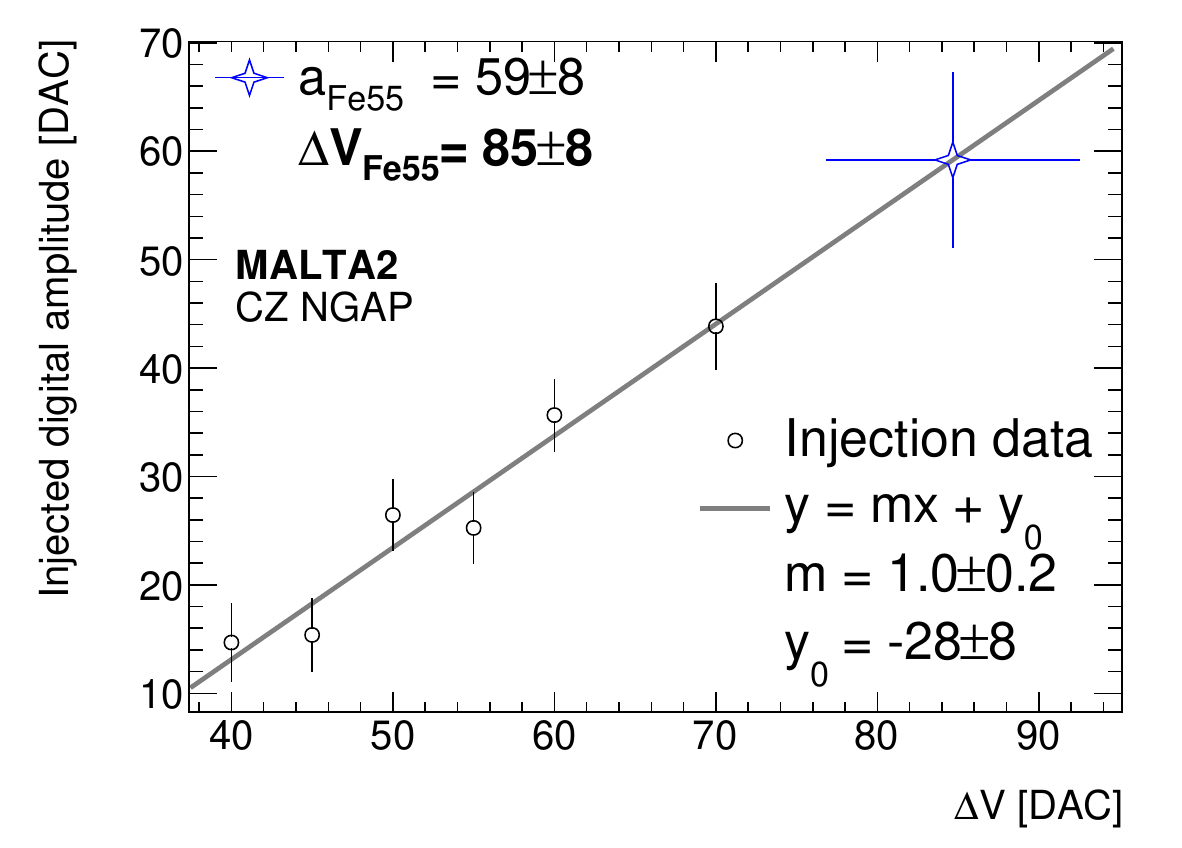}
\caption{W14R11}
\end{subfigure}
\begin{subfigure}[t]{0.49\textwidth}
\centering
\includegraphics[width=\linewidth]{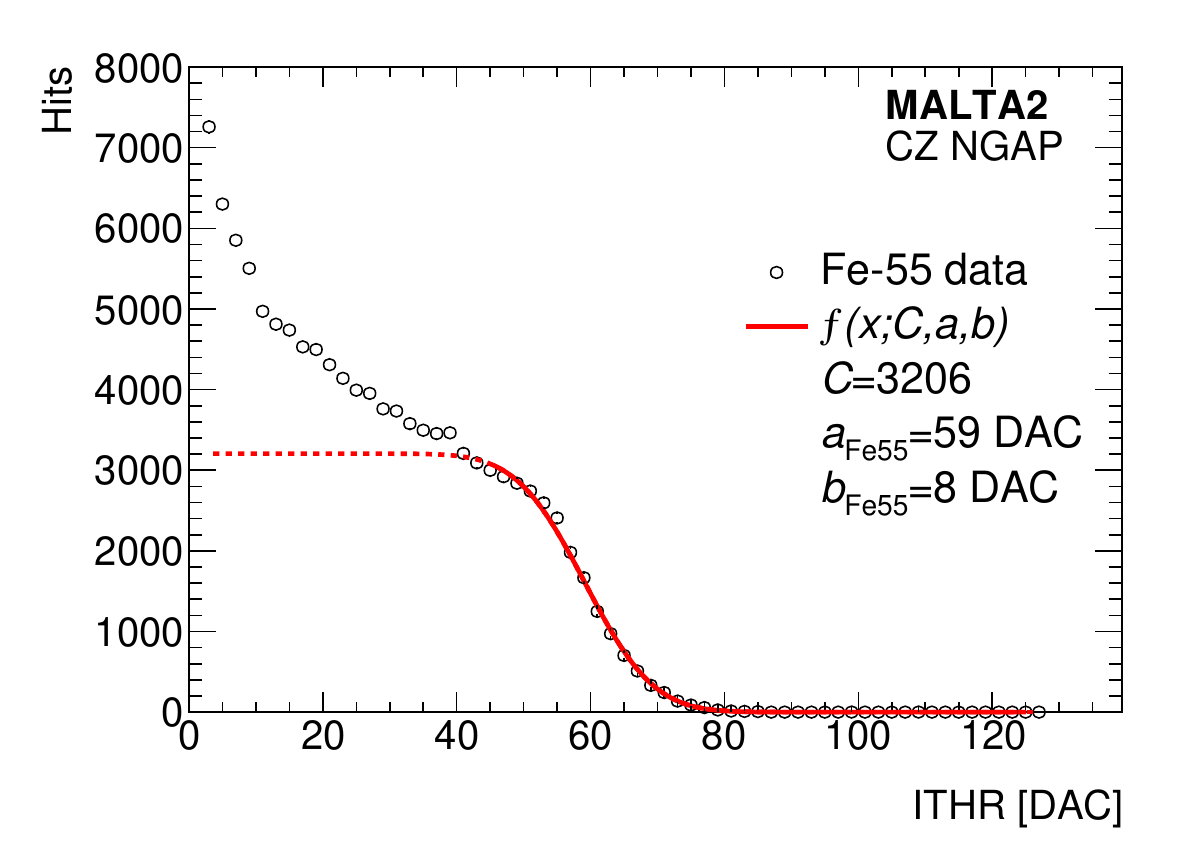}
\caption{W14R11}
\end{subfigure}

\begin{subfigure}[t]{0.49\textwidth}
\centering
\includegraphics[width=\linewidth]{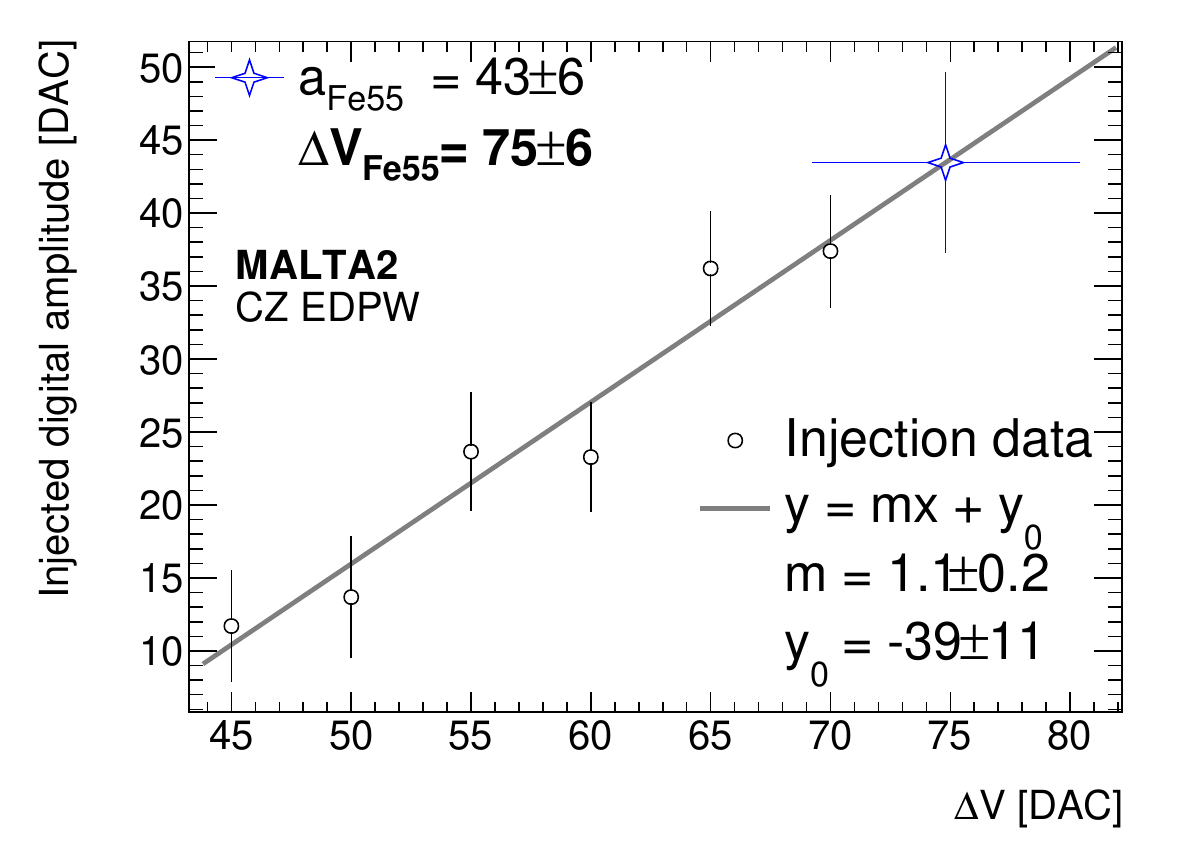}
\caption{W18R17}
\end{subfigure}
\begin{subfigure}[t]{0.49\textwidth}
\centering
\includegraphics[width=\linewidth]{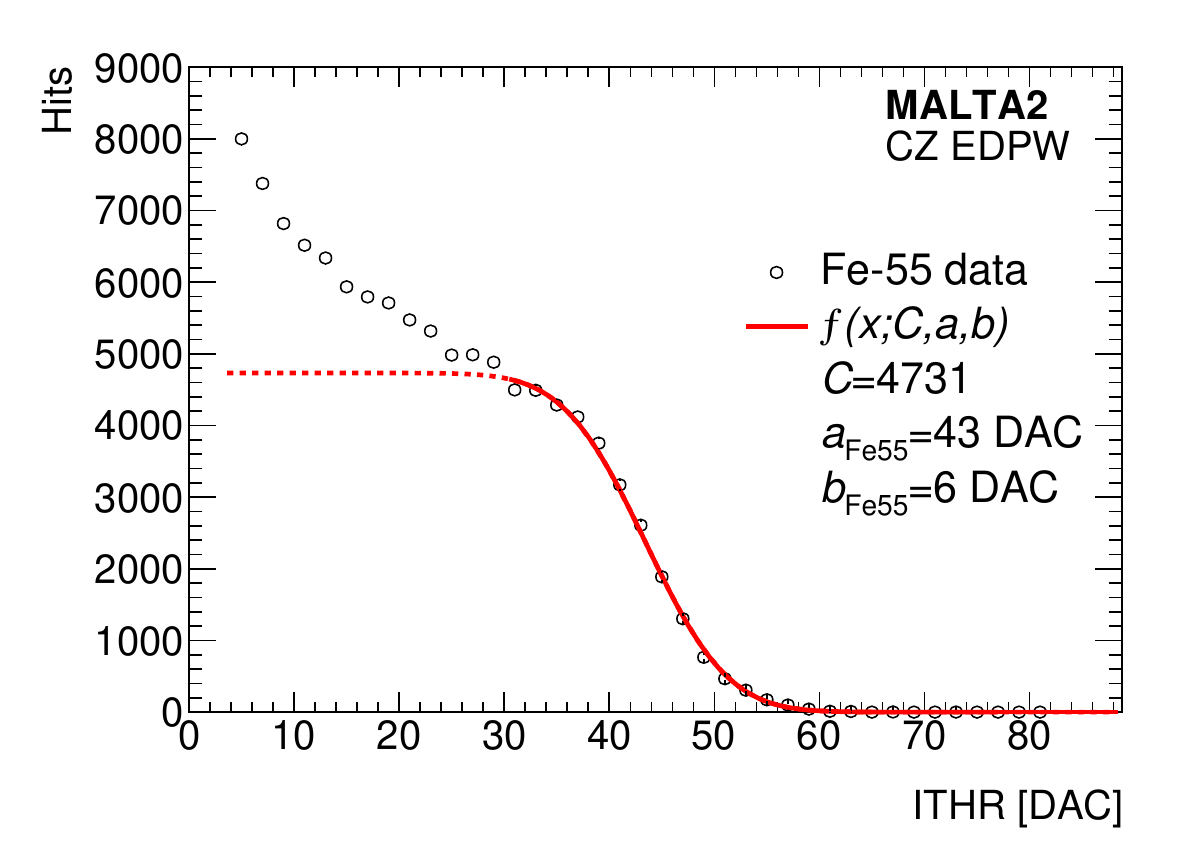}
\caption{W18R17}
\end{subfigure}

\begin{subfigure}[t]{0.49\textwidth}
\centering
\includegraphics[width=\linewidth]{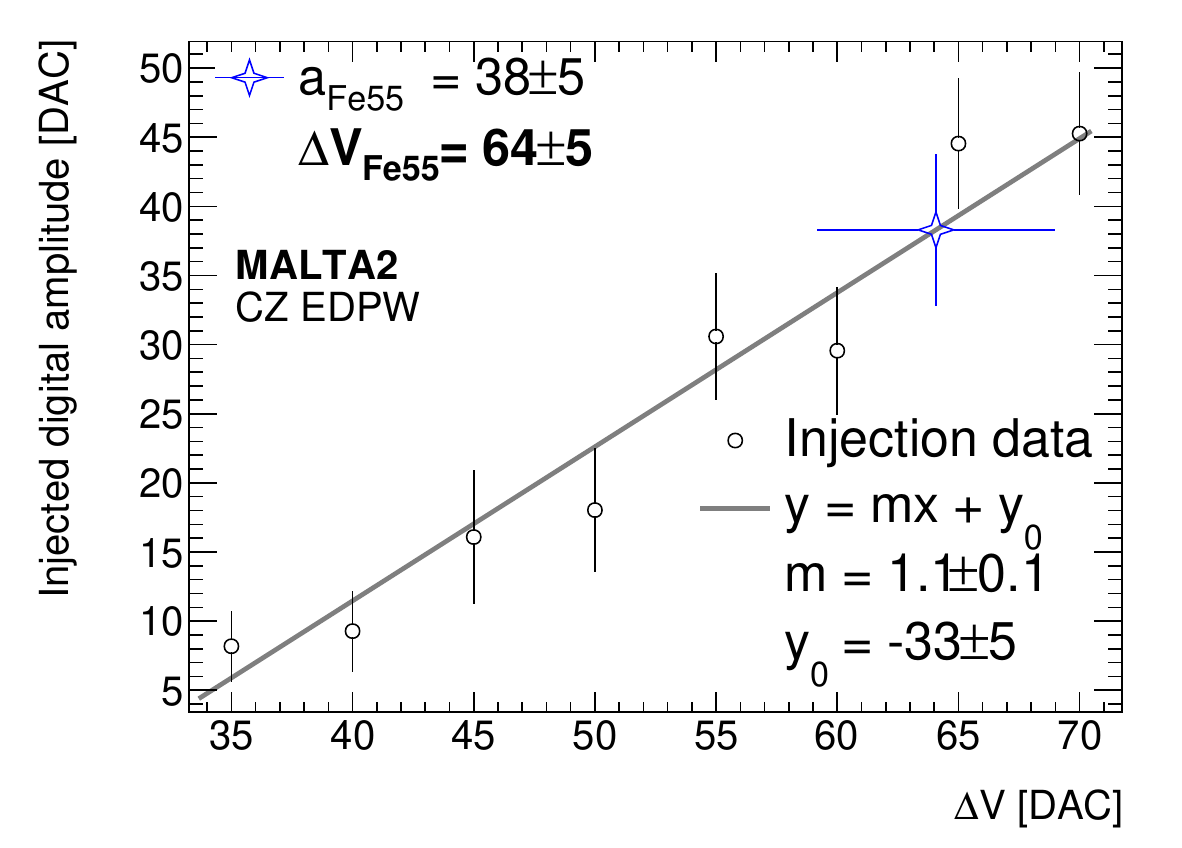}
\caption{W18R19}
\end{subfigure}
\begin{subfigure}[t]{0.49\textwidth}
\centering
\includegraphics[width=\linewidth]{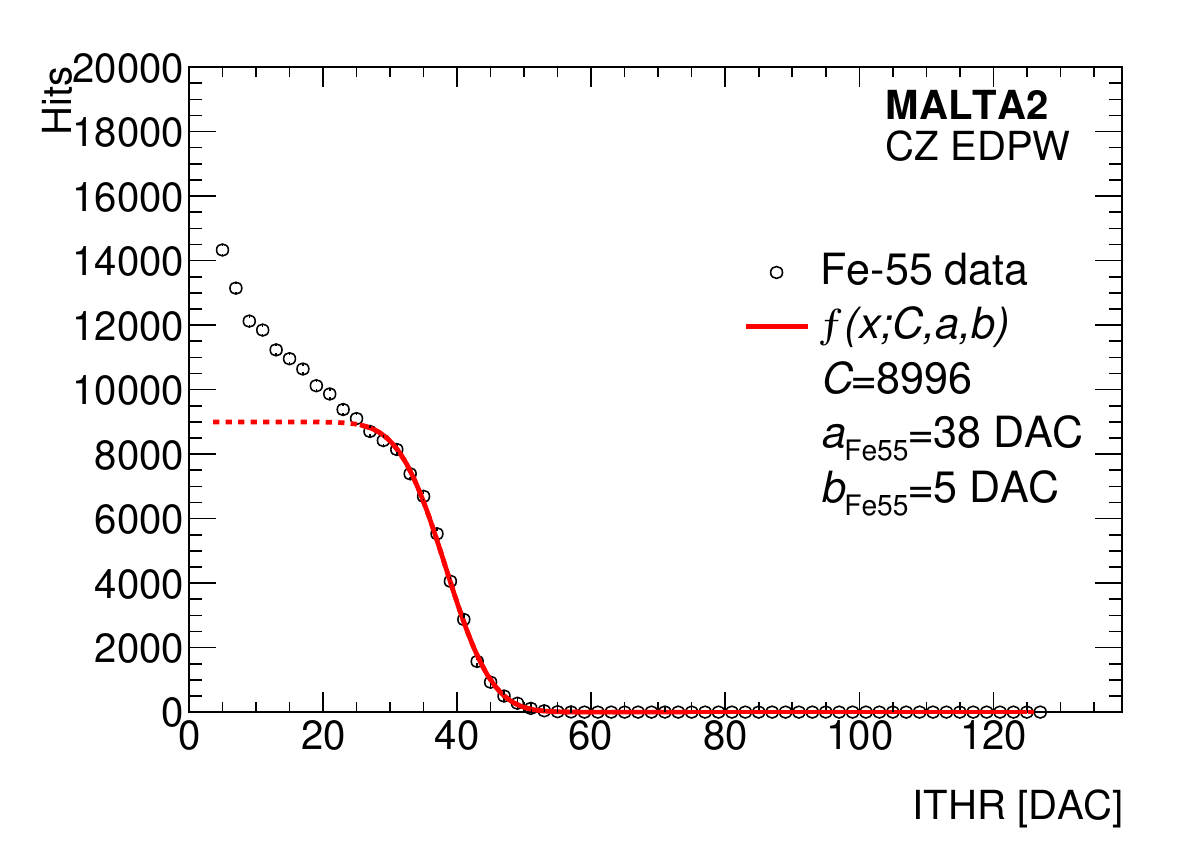}
\caption{W18R19}
\end{subfigure}
\caption{}
\label{fig:Append_Calib2}
\end{figure*}

\begin{figure*}
\centering

\begin{subfigure}[t]{0.49\textwidth}
\centering
\includegraphics[width=\linewidth]{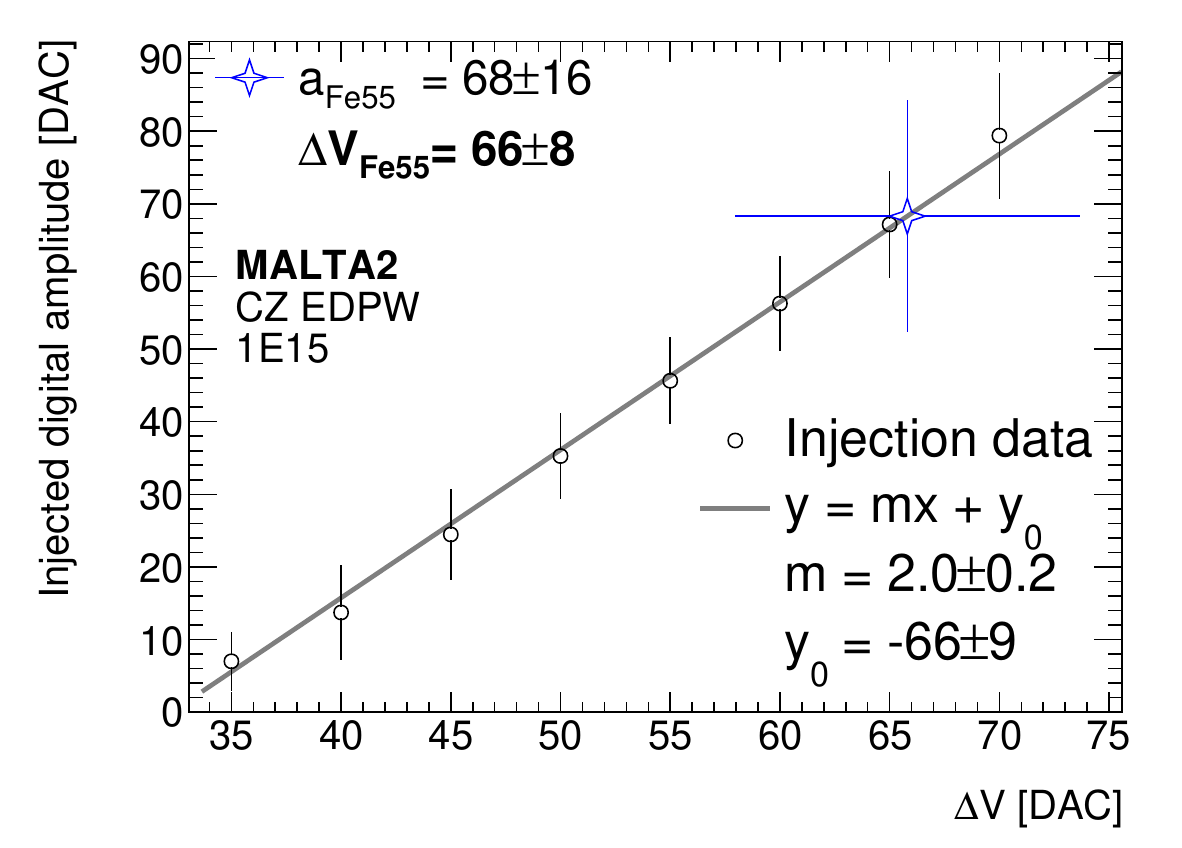}
\caption{W12R7 (1E15)}
\end{subfigure}
\begin{subfigure}[t]{0.49\textwidth}
\centering
\includegraphics[width=\linewidth]{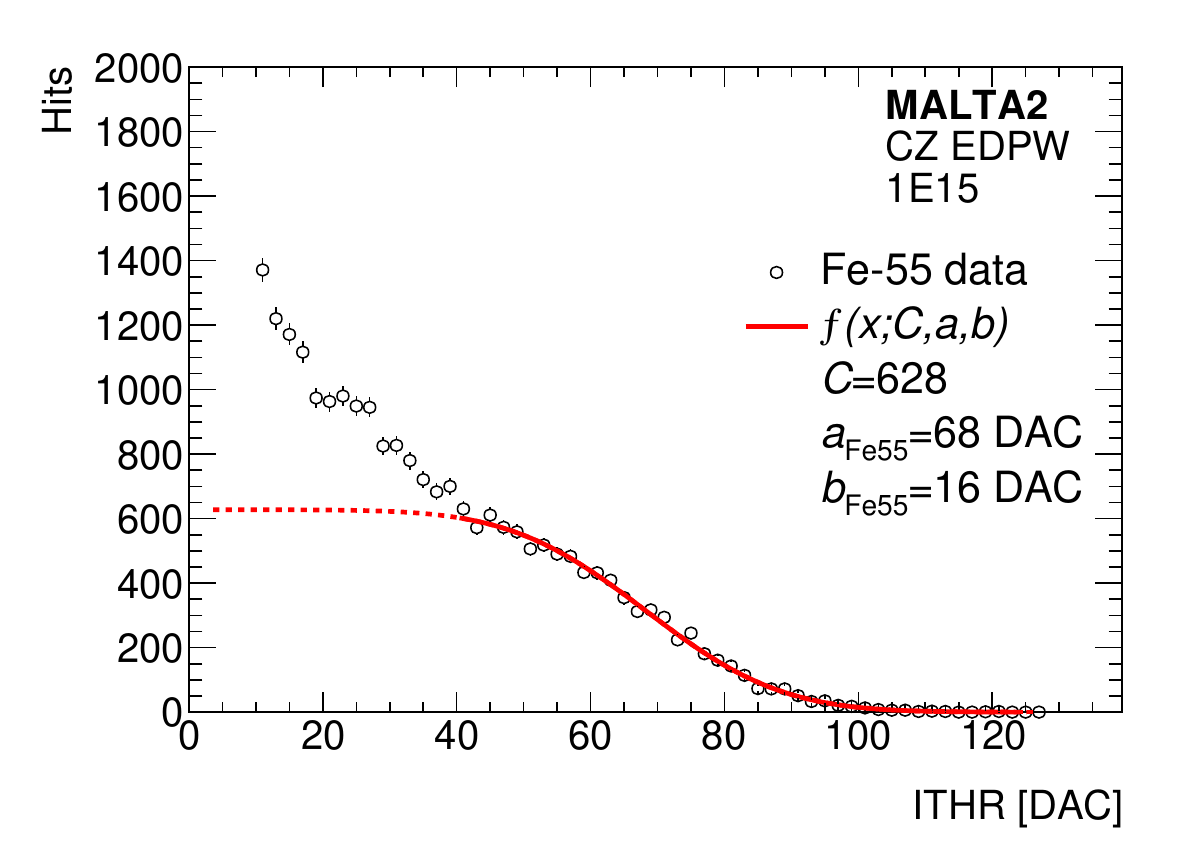}
\caption{W12R7 (1E15)}
\end{subfigure}

\begin{subfigure}[t]{0.49\textwidth}
\centering
\includegraphics[width=\linewidth]{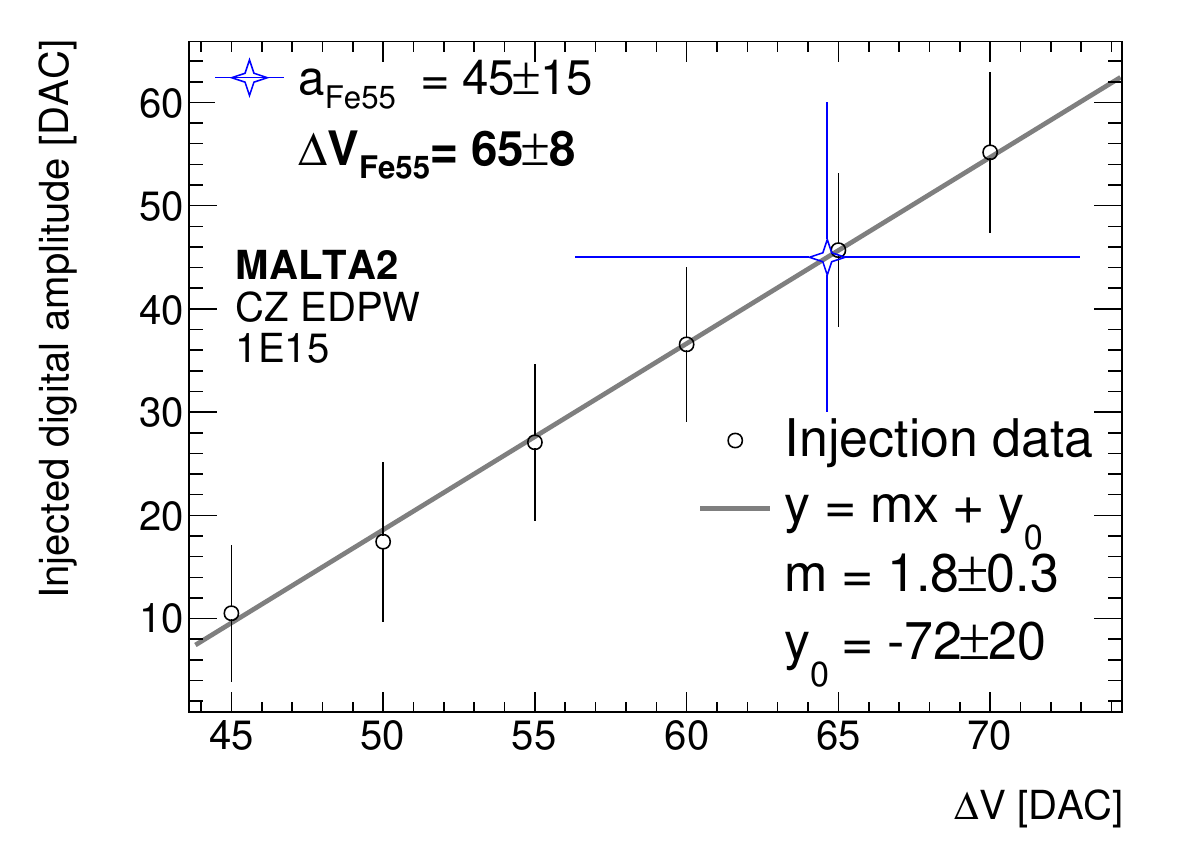}
\caption{W18R1 (1E15)}
\end{subfigure}
\begin{subfigure}[t]{0.49\textwidth}
\centering
\includegraphics[width=\linewidth]{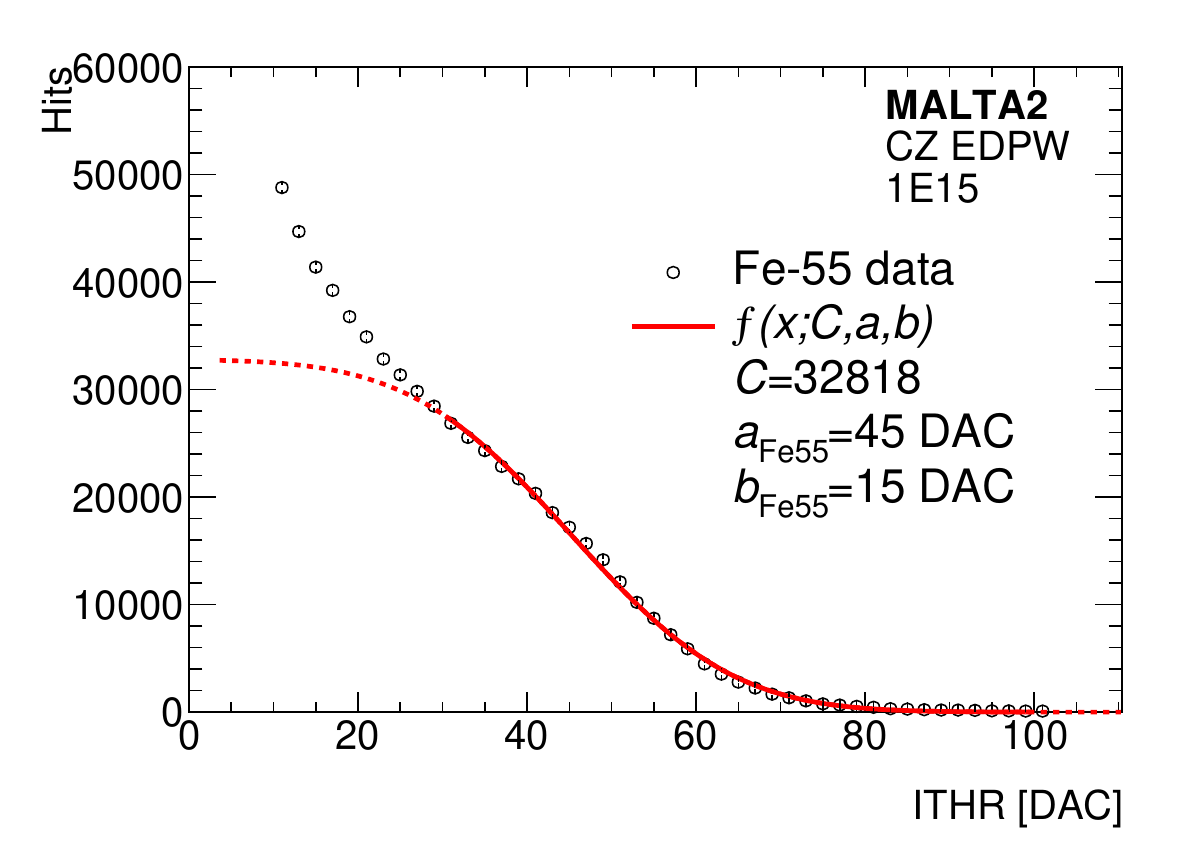}
\caption{W18R1 (1E15)}
\end{subfigure}

\begin{subfigure}[t]{0.49\textwidth}
\centering
\includegraphics[width=\linewidth]{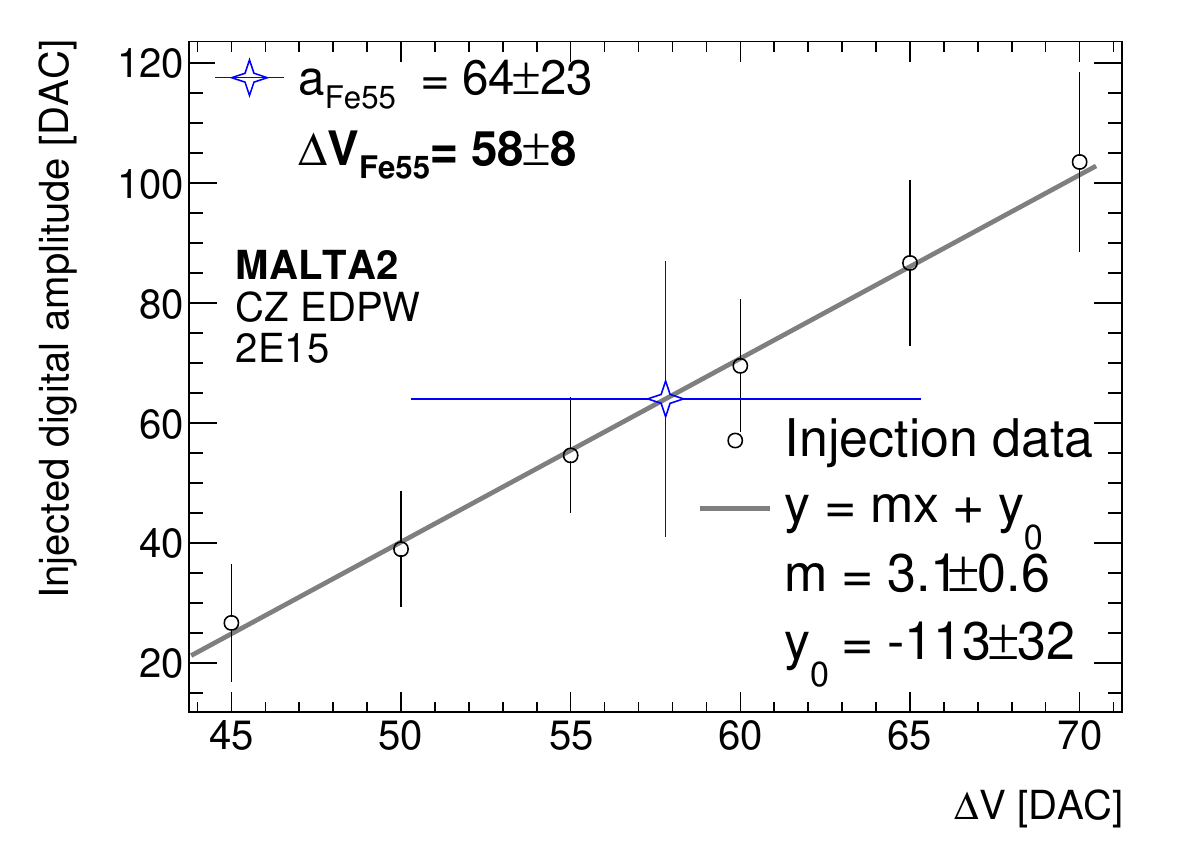}
\caption{W18R4 (2E15)}
\end{subfigure}
\begin{subfigure}[t]{0.49\textwidth}
\centering
\includegraphics[width=\linewidth]{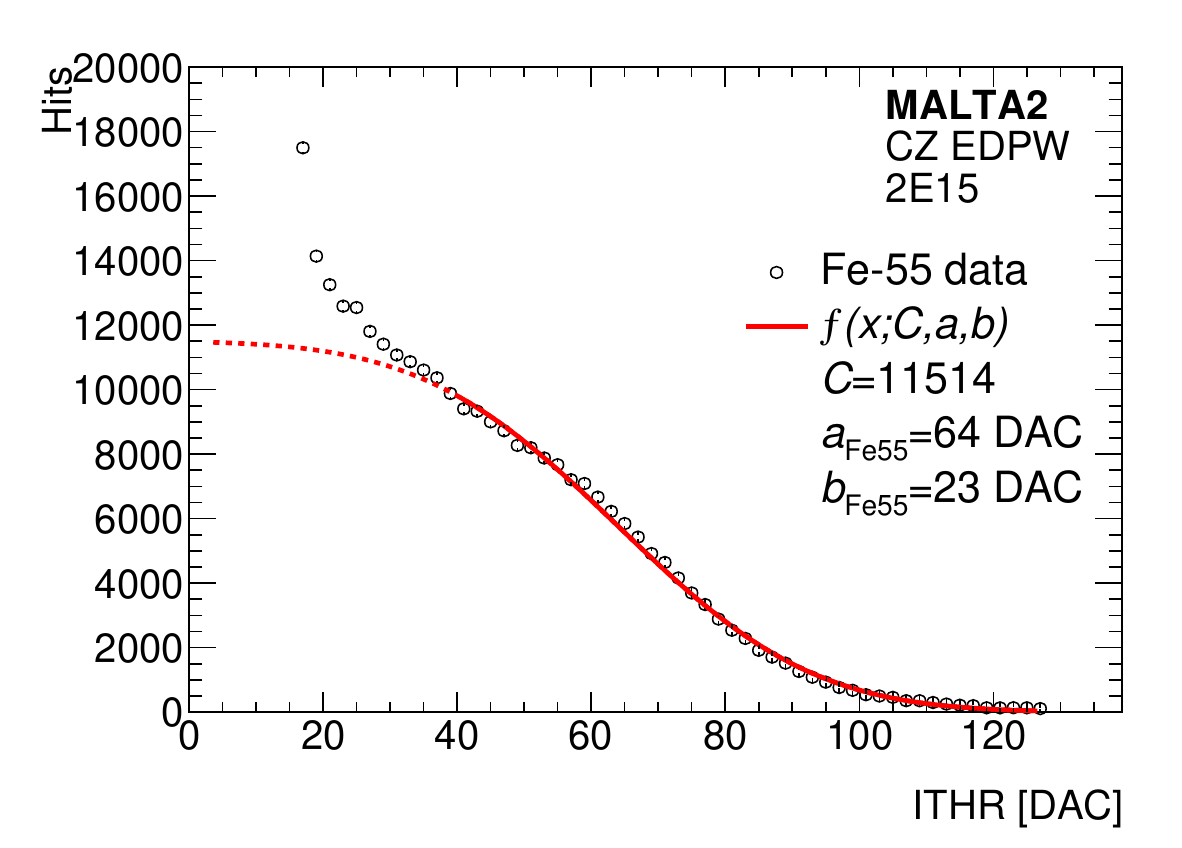}
\caption{W18R4 (2E15)}
\end{subfigure}

\caption{}
\label{fig:Append_Calib3}
\end{figure*}

\begin{figure*}
\centering

\begin{subfigure}[t]{0.49\textwidth}
\centering
\includegraphics[width=\linewidth]{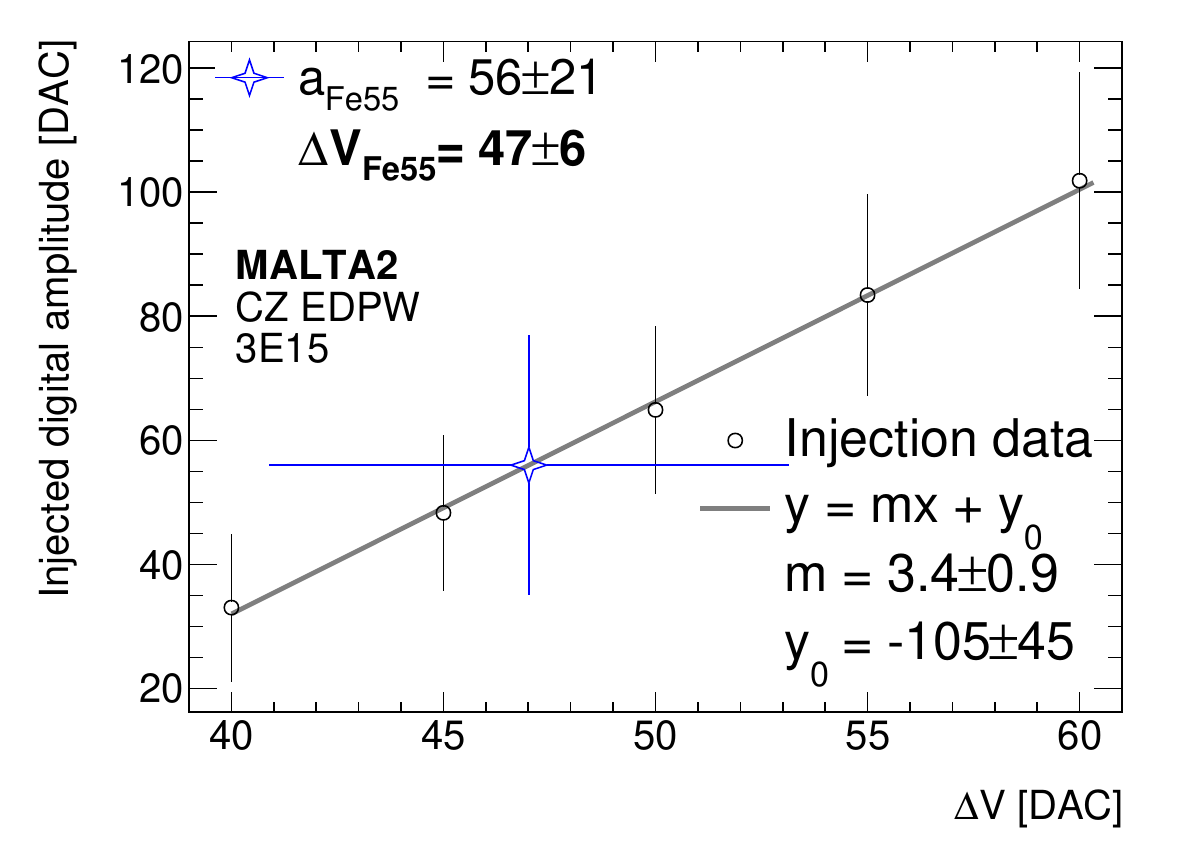}
\caption{W18R9 (3E15)}
\end{subfigure}
\begin{subfigure}[t]{0.49\textwidth}
\centering
\includegraphics[width=\linewidth]{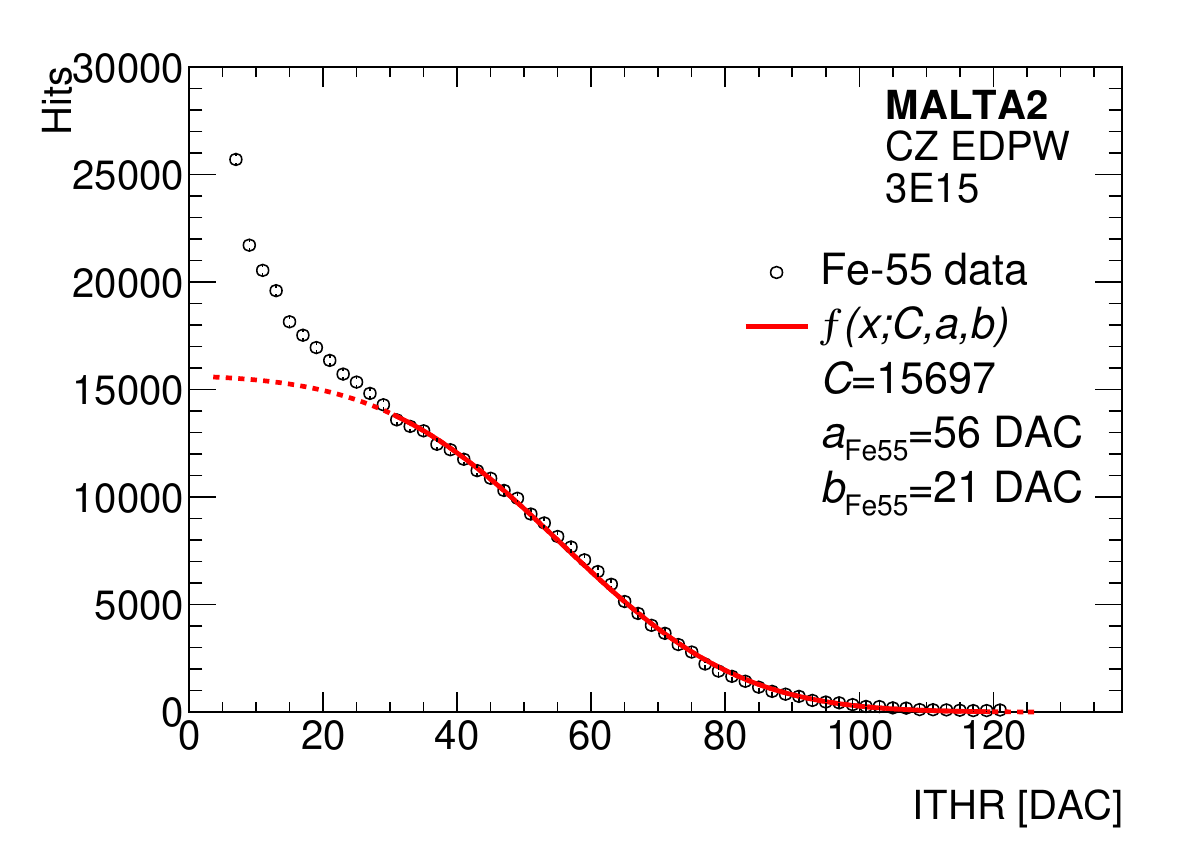}
\caption{W18R9 (3E15)}
\end{subfigure}

\begin{subfigure}[t]{0.49\textwidth}
\centering
\includegraphics[width=\linewidth]{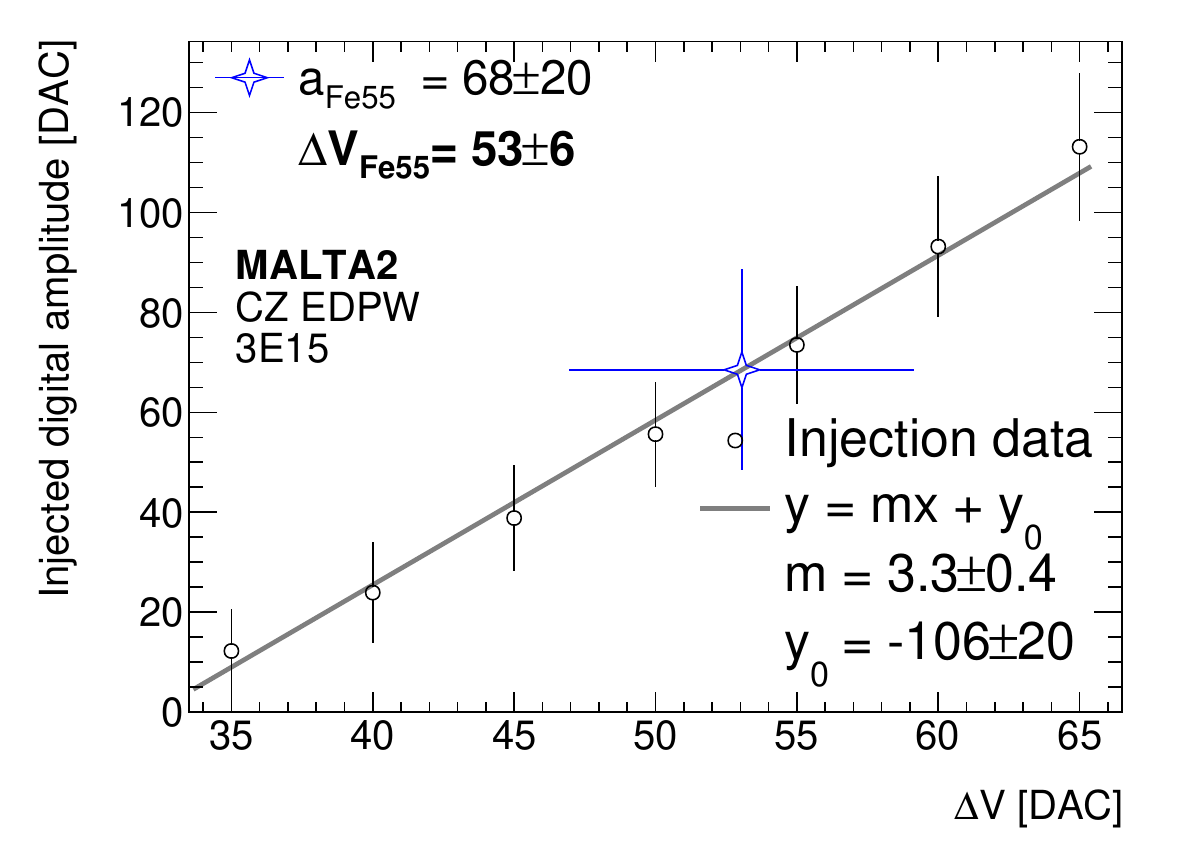}
\caption{W18R21 (3E15)}
\end{subfigure}
\begin{subfigure}[t]{0.49\textwidth}
\centering
\includegraphics[width=\linewidth]{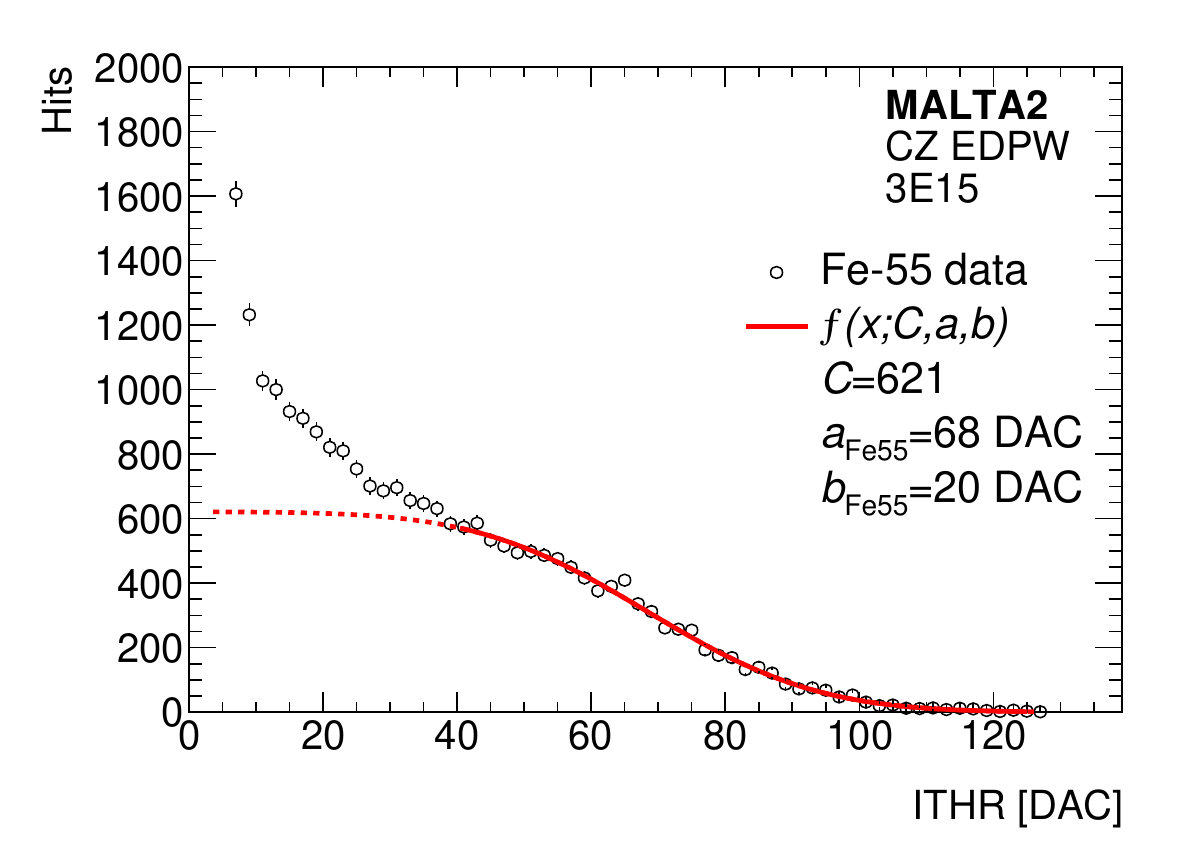}
\caption{W18R21 (3E15)}
\end{subfigure}

\begin{subfigure}[t]{0.49\textwidth}
\centering
\includegraphics[width=\linewidth]{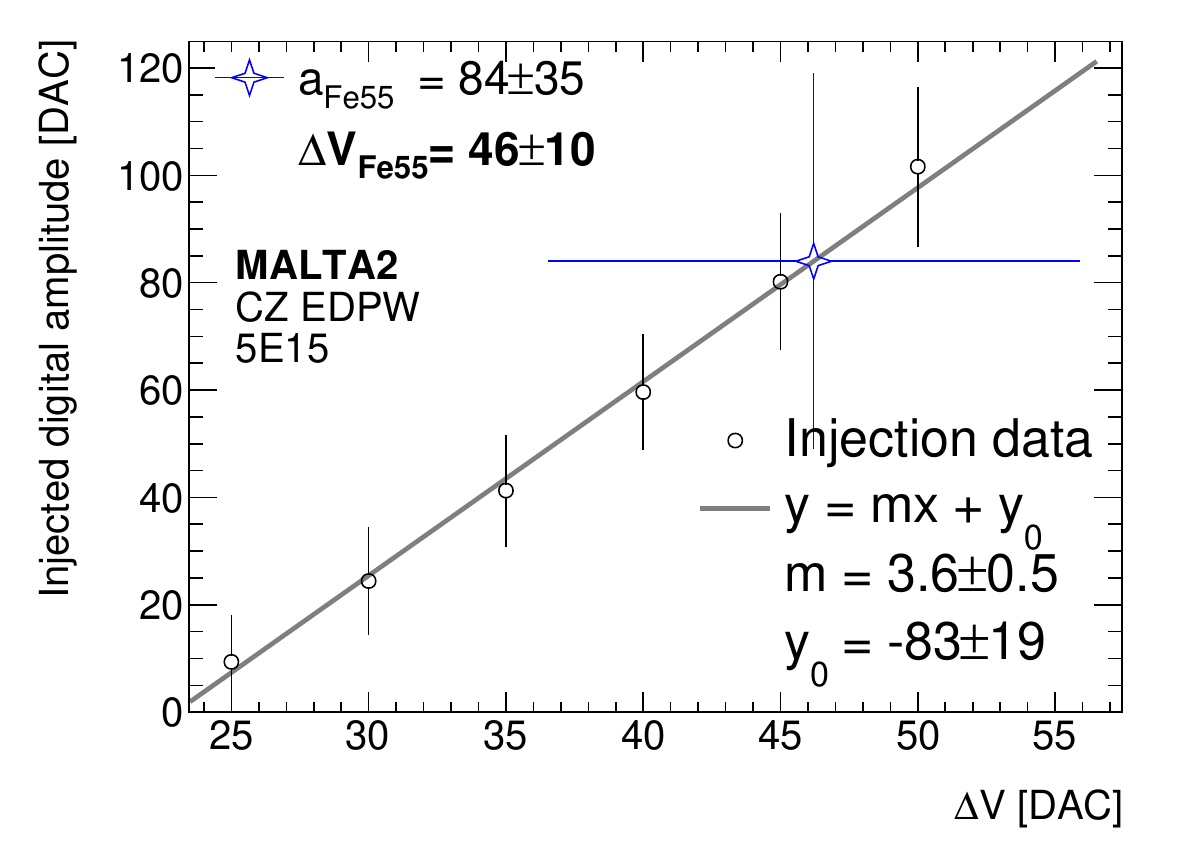}
\caption{W18R12 (5E15)}
\end{subfigure}
\begin{subfigure}[t]{0.49\textwidth}
\centering
\includegraphics[width=\linewidth]{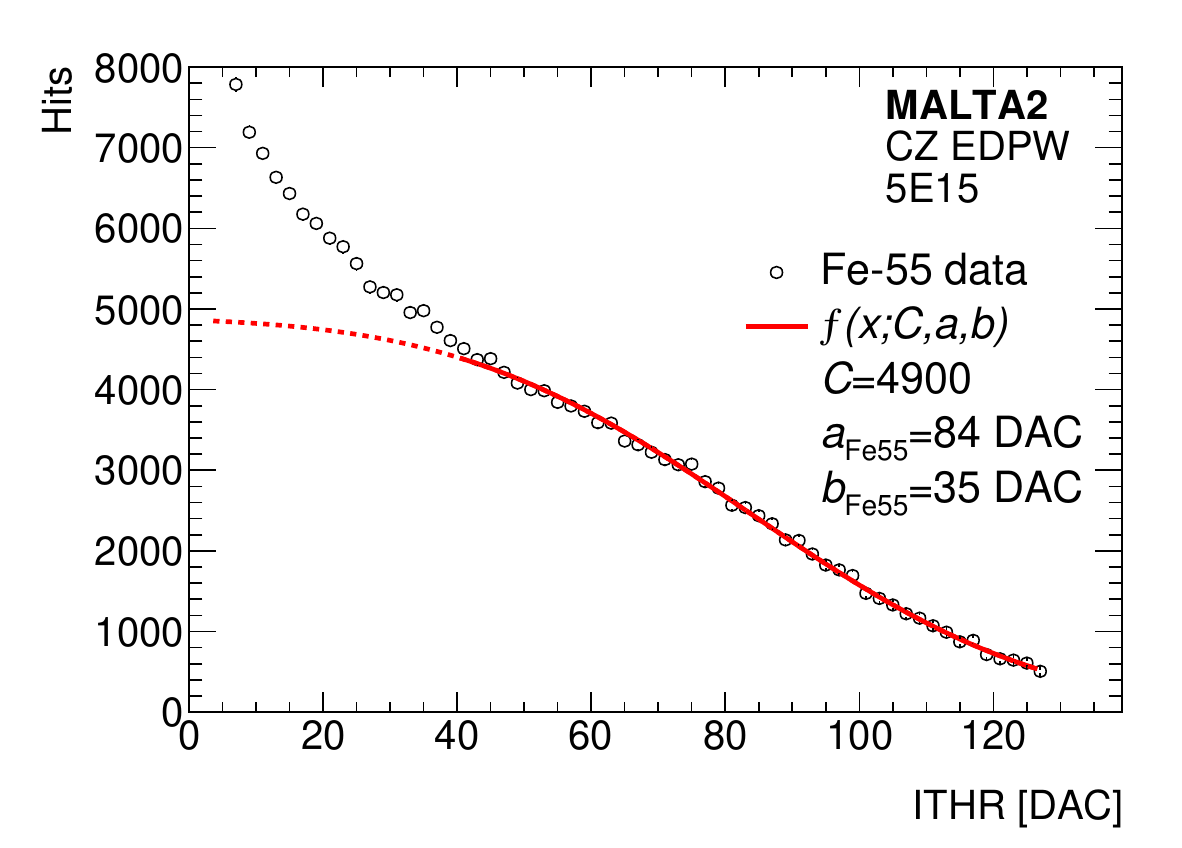}
\caption{W18R12 (5E15)}
\end{subfigure}

\caption{}
\label{fig:Append_Calib4}
\end{figure*}

\begin{figure*}
\centering

\begin{subfigure}[t]{0.49\textwidth}
\centering
\includegraphics[width=\linewidth]{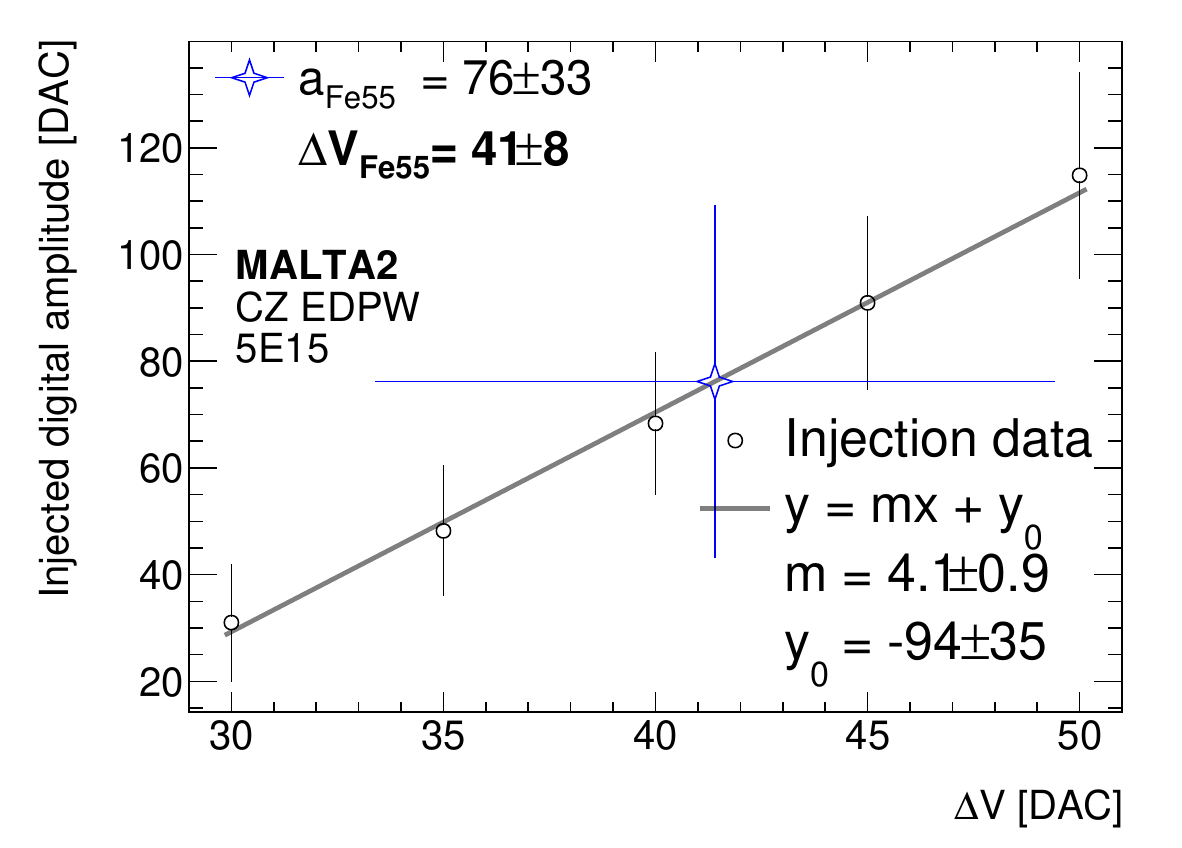}
\caption{W18R14 (5E15)}
\end{subfigure}
\begin{subfigure}[t]{0.49\textwidth}
\centering
\includegraphics[width=\linewidth]{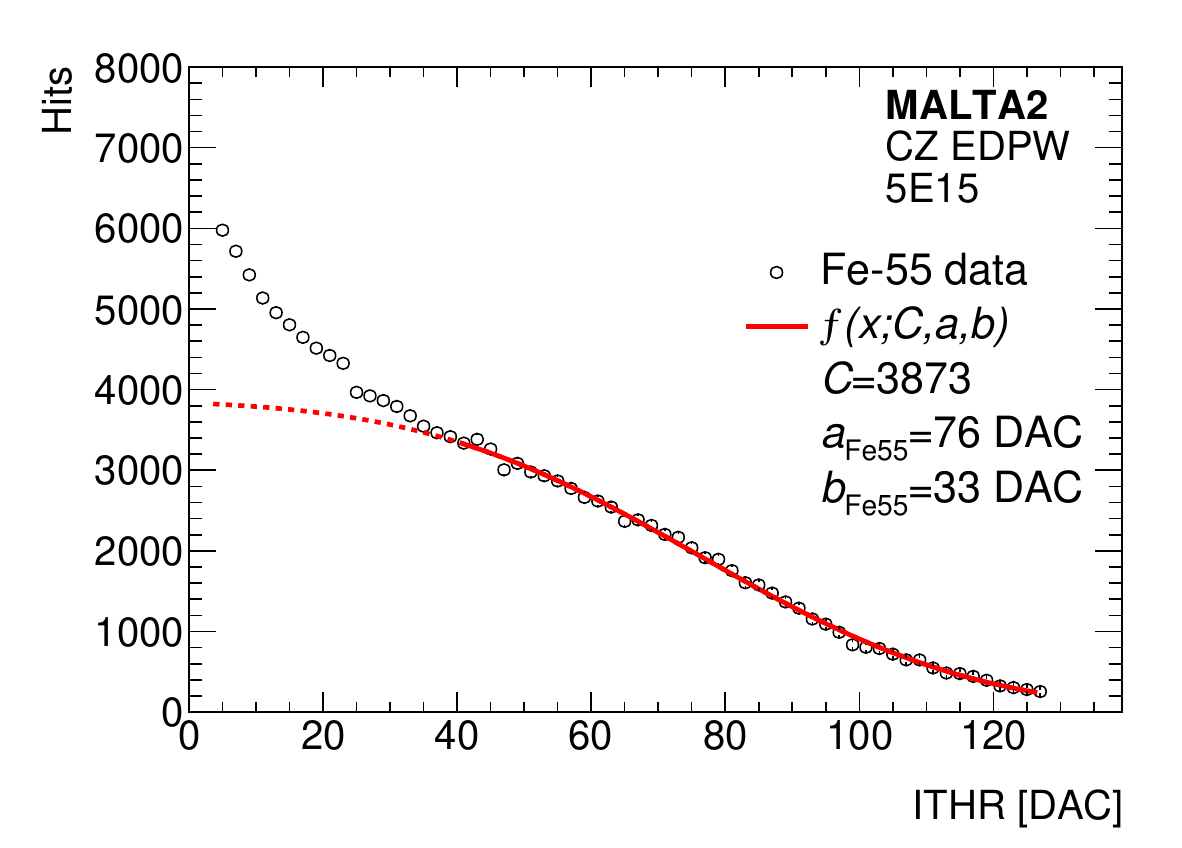}
\caption{W18R14 (5E15)}
\end{subfigure}
\caption{}
\label{fig:Append_Calib5}
\end{figure*}

Figures \ref{fig:Append_Calib1}-\ref{fig:Append_Calib5} show the calibration results for all 13 calibrated sensors of which six are not irradiated. 
They correspond to the Fe-55 threshold scan and injection calibration as described for the example sensor in figs. \ref{fig:Fe55ITHRScan}, \ref{fig:InjectionCalibration}. 
All calibration results are summarized in table \ref{tab:CalibResults}.

\end{appendices}

\end{document}